# The fastest routes of approach to dwarf planet Sedna for study its surface and composition at the close range

Vladislav Zubko[1, 2]

[1]Space Research Institute, ul. Profsoyuznaya 84/32, Moscow, 117997 Russia,

[2]Bauman Moscow State Technical University (National Research University), 2nd Baumanskaya ul. 5, Moscow, 105005 Russia

Current research focuses on designing fast trajectories to the trans-Neptunian object (TNO) (90377) Sedna to study the surface and composition from a close range. Studying Sedna from a close distance can provide unique data about the Solar System evolution process including protoplanetary disc and related mechanisms. The trajectories to Sedna are determined considering flight time and the total characteristic velocity (ΔV) constraints. The time of flight for the analysis was limited to 20 years. The direct flight, the use of gravity assist manoeuvres near Venus, the Earth and the giant planets Jupiter and Neptune, and the flight with the Oberth manoeuvre near the Sun are considered. It is demonstrated that the use of flight scheme with ΔVEGA (ΔV and Earth Gravity Assist manoeuvre) and Jupiter-Neptune gravity assist leads to the lowest cost of ΔV≈6.13 km/s for launch in 2041. The maximum payload for schemes with ΔVEGA manoeuvre is 500 kg using Soyuz 2.1.b, 2,000 kg using Proton-M and Delta IV Heavy and exceeds 12,000 kg using SLS. For schemes with only Jupiter gravity assist, payload mass is twice less than for ones with ΔVEGA manoeuvre. As a possible expansion of the mission to Sedna, it is proposed to send a small spacecraft to another TNO during the primary flight to Sedna. Five TNOs suitable for this scenario are found, three extreme TNOs 2012 VP113, (541132) Leleākūhonua (former 2015 TG387), 2013 SY99) and two classical KBOs: (90482) Orcus, (20000) Varuna

## 1. Introduction

### 1.1 Astrophysical characterisation of Sedna

Icy bodies placed beyond the orbits of the giant planets are of tremendous interest for studying processes that took place in the earliest stages of the evolution the Solar System. Some of these bodies (centaurs) are located between the orbits of Jupiter and Neptune. The others are called trans-Neptunian objects (TNOs) because their orbits are beyond the orbit of Neptune. They are classified, according to series of papers [1–5], as objects of the Edgeworth-Kuiper Belt (KBOs) if the semi-major axis (a) >35 AU, the detached disk (a >100 AU) and the hypothetical Oort Cloud (inner part a >500 AU, and outer part a >10,000 AU). The close-range survey of such bodies should enhance the knowledge of the processes that accompanied the early stages of the Solar System's evolution. Our work, though, focuses on the object that stands out from the rest of TNOs.

TNO (90377) Sedna is a small celestial body whose orbit is located far outside the Edgeworth-Kuiper Belt. Its belonging is a subject of much discussion; because of the remoteness of Sedna, it possibly can belong to the inner Oort Cloud as stated by discoverers [6,7]. Still, the International Astronomical Union (IAU) organization, the Minor Planet Center (MPC) and Small[1] assign Sedna to the scattered disc objects. Observations of Sedna since its discovery in 2004 [1,3,5,8–11] indicate that the object is similar in composition to Pluto and classical KBOs. Such estimates are the reason for the interest in the Sedna because, despite the similarities, its orbit is radically different from most KBOs. In aphelion, it moves away from the Sun at a distance exceeding 1,000 AU and never approaches the Sun nearer than 74 AU at perihelion. Such orbital motion cannot be explained, for example, only by resonance with Neptune caused by the Kozai-Lidov effect as it happens with Pluto and similar objects [8,12]. Nowadays, several celestial bodies have been discovered whose orbits are close to Sedna (sednoids). One of these bodies is, for example, the sednoid 2012 VP$_{113}$ (unofficially named as 'Biden') [4]. The presence of similar orbits of numerous TNOs gives rise to different theories about their origin; the most popular is the theory of scattering objects by the gravity of a massive celestial body, the so-called Ninth Planet hypothesis [13,14]. The ideas of the origin of Sedna are diverse. The discoverers [8] have suggested that Sedna was created in the Solar System at the early stage of its evolution, and its orbit was changed because of the dynamic effects that followed the Sun's formation within a dense stellar cluster [3,8,15–17]. According to other versions, Sedna's orbit was changed by a stellar encounter [4] (e.g., the passing Scholz's star about 70

---

[1] The IAU MPC official web page. Url: https://minorplanetcenter.net/data (Accessed September 17, 2021)

thousand years ago [18] at a distance of 52,000 AU from the Sun), or Sedna had been captured from a low-mass star or a brown dwarf in interstellar space [19].

As of today, observations of Sedna have been carried out only with the help of ground and near-Earth orbital instruments. Observations at wavelengths of 0.4 - 2.4 μm [9,11] as well as over 2.5 μm [9,11] have shown that the surface of Sedna is abundant in ice (>50% [9]) and consists of a mixture of $CH_4$, tholin layer, $NH_3$, $N_2$ and $H_2O$. The assumption of water ice on the surface of Sedna was made from infrared observations using the Infrared Array Camera (IRAC) on the Spitzer Space Telescope at wavelengths above 2.5 μm. The presence of $N_2$ emission lines in the near-infrared spectrum may indicate the presence of a thin atmosphere in this dwarf planet. Still, such an atmosphere may exist when Sedna moves near the Sun, i.e. ~200 yrs out of a 10,500-year orbital period. Paper [20] refers to the possible existence of a subsurface ocean inside Sedna. This theory is explained by the presence of ammonia in concentrations up to 1.4% and the estimated size of the object is from 650 to 1,500 km [9,10,20,21]. The presence of ammonia makes it possible to reduce the melting point of water ice inside Sedna. The object's size makes it possible to judge that at a depth of about 227 km, a 14 km layer of liquid water is warmed by the heat of internal radioactive decay, e.g. $^{26}Al$.

The presence of tholins on the surface of Sedna is probably the reason for its bright red colour observed with twin Magellan Baade and Clay 6.5 m telescopes at Las Campanas, Chile and the 8.2 m Subaru telescope atop Mauna Kea in Hawaii [1,9]. In particular, research [1] shows that such colouring is inherent to most Sedna-like objects with aphelion more than 500 AU. According to the same source, the reason for it may be that sednoids stay outside the heliosphere most of the time and are not exposed to the thermal radiation of the Sun. At the same time, the surfaces of long-period comets with a perihelion of the order of 10 AU and an aphelion several orders of magnitude larger than that of Sedna, observed with telescopes, are either blue or red with a spectral gradient (S) <10 (note, the S of Sedna is estimated at ~25 units) [1]. This difference in the colouring of the surface of the objects, presumably all having the same origin, may be explained by the intense heating of the surface by the thermal radiation of the Sun during the close approach [1,9–11]. It should also be noted that cosmic rays irradiate Sedna's surface. Therefore, it is interesting to study the interaction of organic compounds and water ice with cosmic rays in the interstellar medium.

The study of the object at close range might considerably enhance the results obtained previously and provide unique data about the early evolution of the Solar System, particularly on the origin of the protoplanetary disk and the mechanisms of its existence.

*1.2 Orbit design to Sedna*

Researches toward studies of a flight to Sedna began since 2011. The paper [22] defines the best scenario of reaching Sedna with a single Jupiter gravity assist (JGA). According to [22], such a flight would require a total cost of 7.42 km/s of characteristic velocity (ΔV) and a time of flight (TOF) of 24.48 yrs. It is worth highlighting the studies carried out in the paper [23], which considers flight scenarios to Varuna, Haumea, Makemake, Ixion and Sedna. The research carried out in papers [23,24] showed that increasing ΔV and using JGA makes it possible to reach Sedna in 18-20 years of TOF. However, their estimate of necessary ΔV was about 14.5 km/s.

A flight to Sedna for the subsequent insertion into the orbit of its satellite would require both a large launch ΔV and a braking impulse because the mass of Sedna is small and the velocity of the spacecraft, acquired through gravity assist manoeuvres, is too high. Estimates of orbiting Sedna using high thrust are given in [23]. Therefore, the only way to study Sedna becomes its observation from a flyby trajectory. Note, the authors of the paper [24] considered the flight to Sedna with electric propulsion (low thrust) and braking to a satellite orbit of Sedna. According to them, the orbiting Sedna is possible, using low thrust, and the mass of the satellite can be about 50-100 kg if heavy or super-heavy launch vehicle is used [24].

Only one mission in history is known to have successfully explored TNOs at close range. New Horizons successfully explored Pluto and the TNOs (486958) Arrokot from a flyby trajectory. Based on this mission data, several good objectives for the study of Sedna can be highlighted [25–28], as follows:

- Sedna surface spectrometry;
- the presence of a magnetic field of the object;
- investigation of the effect of solar wind and interstellar radiation on organic compounds (tholins), water ice on the surface of the object;
- study of the presence of an atmosphere;
- testing the hypothesis of a subsurface ocean (via impact mission/using descent probes or capsules, or ~~a~~ sending the spacecraft to the near Sedna orbit);
- surface composition study by taking a sample of the Sedna soil ejected by means of hitting the surface by a projectile separated from the spacecraft before the Sedna encounter;
- clarification of the mass of the object [29];

- obtaining images of Sedna in the optical range, conducting photo and video study sessions of the object;
- studies of the dust composition of the Kuiper Belt;
- specific studies of the solar wind at long distances from the Sun.

The papers [30–38], which consider scenarios of flights to other TNOs and, in particular, to the interstellar body 1I/Oumuamua [30–32], are also of interest. The authors of [35] analyse the application of tether systems for a flight to Haumea and other Edgeworth-Kuiper Belt objects. In [23], the use of the JGA and other gravity assists for the flight to 45 TNOs are analysed. In their work on the flight to the 1I/Oumuamua asteroid, the authors examine the use of predominantly flight schemes, including the Oberth manoeuvre near the Sun [31,39]. Due to the close approach to the Sun, a small impulse in the vicinity of the perihelion allows to significantly increase the heliocentric velocity of the spacecraft [39]. The results show that it is possible to reach 1I/Oumuamua at a distance of 70 to 110 AU with such an approach. The cost of such flight will be from 15 to 30 km/s within TOF from 18 to 22 yrs correspondingly [31]. Research [39–42] of interplanetary and interstellar flight using laser solar sail and direct fusion drive is worth mentioning. The results given in [41] show that, for example, a flight to Haumea, Makemake and Eris will only take 6 to 10 yrs at a propellant consumption of about 3.5-4 tonnes.

In the works [43,44], trajectories to Sedna were determined satisfying two constraints simultaneously, namely, with the TOF no longer than 50 yrs, the total characteristic velocity ($\Delta V_\Sigma$ that is the sum of the characteristic velocity required for launch, as well as all manoeuvres in deep space and near planets) had to be less than 8 km/s. In all cases, the correcting manoeuvres were not considered since they depend on the navigation and the manoeuvres execution accuracy. Thus, the problem of the flight to Sedna was reduced to the search of the minimum $\Delta V_\Sigma$ under the restrictions on the TOF.

The current work focuses on the search for flight schemes to Sedna, ensuring the fastest way to reach it under the only restriction on $\Delta V_\Sigma$ being the natural limitation of the chemical propulsion systems. Because of this, flights restricted on duration to 10-20-yr are considered, that is admissible because the average time of satellite's lifetime is evaluated approximately in 15 yrs. On the other hand, all previous missions to deep space such as Pioneer 10, Pioneer 11, Voyager 1, Voyager 2, Galileo, Ulysses, New Horizons and others lasted at least 5-10 yrs before they fulfilled their main goal of the mission. Also, the additional problem that is to find the TOF which corresponds to the limited $\Delta V_\Sigma$ was considered in this paper. Note that in all flight schemes to Sedna that will be analysed in this paper the costs of orbit correction manoeuvres are not considered, as well as in the paper [43].

Note that the trajectory analysis was done only for high thrust, because the main objective of this paper is to demonstrate that it is possible to reach Sedna by well-known means, even when the time of flight is severely limited. In addition, low thrust has a significant disadvantage seen in a flight to such a distant object as Sedna. Namely, since the effectiveness of the thrust decreases with increasing the distance from the Sun (correct for solar electric propulsion (SEP)), using low thrust beyond Jupiter's orbit does not seem to be a feasible option. In such a case, the use of a fusion engine, a nuclear power plant or a laser solar sail may be considered for a flight to Sedna, but this subject requires special consideration.

This research analyses direct flight in order to compare it with more complex schemes involving gravity assist manoeuvres. Particular attention is given to flight schemes with active manoeuvres (i.e. extra $\Delta V$ costs are needed) in deep space and scheme with an Oberth manoeuvre (OM) near the Sun. Strategies with gravity assist manoeuvres include ones at both Venus and Earth and giant planets. Approximate estimates of the payload mass delivered to Sedna are given.

A possible expansion of the mission to Sedna has also been proposed. In such a scenario, a flight takes place simultaneously to both Sedna and another TNO by separation of a small probe from the spacecraft. Such separation takes place during the Jupiter or Neptune flyby. After that the probe is directed to another TNO, the flight to which is possible without additional manoeuvres in the planetary sphere of influence (SOI). Results are given for five TNOs suitable for this scenario, including three extreme TNOs 2012 VP$_{113}$, (541132) Leleākūhonua (former 2015 TG$_{387}$) and 2013 SY$_{99}$ and two classical KBOs: (90482) Orcus, (20000) Varuna.

**2. Direct flight to Sedna**

First, it is essential to study a direct flight to the object since such a flight has the most remarkable simplicity in terms of orbit construction and control of the spacecraft as it moves towards the target. Also, a direct flight is essential for determining the basic energy characteristics of the trajectories, namely the required cost $\Delta V$ depending on the launch and arrival dates. The value of the required $\Delta V$ is determined by solving the Lambert problem. The point of this boundary value problem is to define the spacecraft's orbit in the framework of the Kepler motion model according to two given positions and the time of spacecraft flight between these positions. There are many methods for solving this problem; in this paper, we use [45,46], which are based on a universal variable.

To analyse the direct flight, we plot the ΔV vs launch and arrival dates for flight to Sedna (Fig. 1).

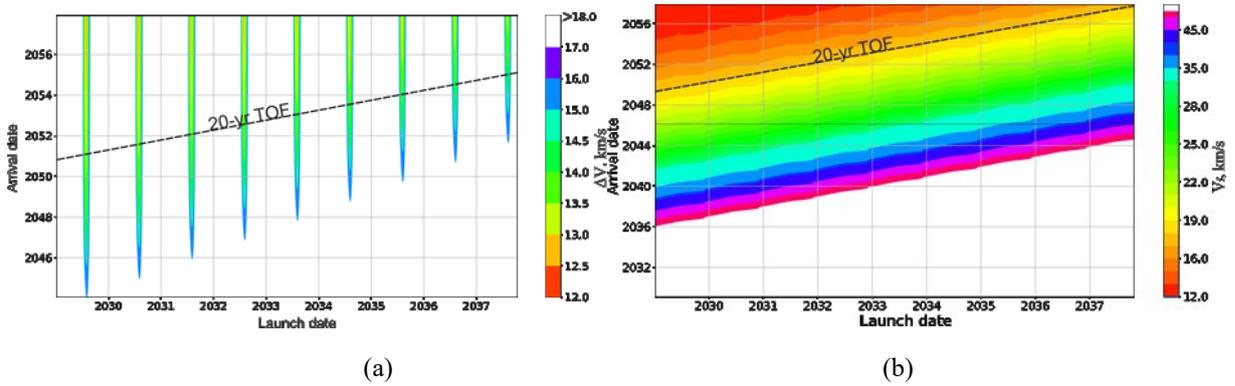

(a) (b)

Fig. 1 (a) - ΔV vs launch and arrival dates to Sedna, (b) – spacecraft's rendezvous velocity with Sedna ($V_f$) vs launch and arrival dates to Sedna.

The ΔV value (Fig. 1a) show that its value for the flight to Sedna cannot be less than 14.5 km/s for the TOF less than 20-yr. This value of the launch ΔV ($ΔV_0$ in this section equal to ΔV) for a flight to Sedna will require to use super-heavy launch vehicle with at least 3-4 additional stages.

Note that the value of the spacecraft velocity of Sedna flyby ($V_f$) (Fig. 2b) is exceeds 15 km/s for a 20-yr flight what is too high. Since the gravitational parameter of Sedna is small, the impulse required for the spacecraft to enter the orbit of the satellite of Sedna will be approximately equal to the relative velocity of the spacecraft. It means that the total cost for launch and breaking to the orbit of Sedna will be at least 29 km/s, which is currently quite fantastic, rather than a realistic value of the characteristic velocity when planning a direct mission with a final insertion to the orbit Sedna satellite.

### 3. The fastest routes to Sedna using gravity assist manoeuvres

#### 3.1 Methods of trajectories calculation

In this work, the patched conic approximation [47–51] is used as a model of spacecraft motion. In this approximation, it is supposed that during the spacecraft flight in the planets' SOI, the spacecraft motion is affected only by the gravity force of the central body and beyond the SOI (i.e. on the heliocentric arc) influenced only by the gravity force of the Sun. In other words, all perturbing forces, including solar radiation pressure, oblateness, attraction of other planets on the spacecraft motion, are neglected. While the spacecraft moves inside the SOI, its size is considered infinite, i.e. the planetocentric velocity of the spacecraft on the SOI boundary is assumed to be equal to its asymptotic velocity. As for the heliocentric trajectories, the planetary SOI are considered point ones permissible at the first step of trajectories design considering the size of the planets SOI compared to heliocentric distances. Therefore, the n-body problem is reduced to a set of two-body problems; a solution of each of them is a Keplerian orbit (i.e., a conic section). The problem is to find these orbits and to patch them into a whole trajectory [52].

The parameters of arrival and departure heliocentric arcs are determined using the Lambert problem solution. In the next step, they are connected in their pericenters, so that their heights are equal and tangents coincide. The magnitude of the difference of the velocity vectors at the pericentre of these hyperbolas is equal to the impulse ΔV needed to connect the trajectory's arrival and departure arcs. If the planet's gravitational field is incapable of turning the asymptotic velocity vector from its incoming direction to the outgoing one, an additional impulse rotating the velocity vector to the necessary direction is required.

A two-step procedure is used to search for optimal trajectories, i.e. trajectories that ensure a minimum $ΔV_Σ$ under the given constraints. In the first step, the minimum is calculated using a differential evolution (i.e. a part of genetic algorithms family). The differential evolution belongs to the family of metaheuristic algorithms of the extremum search [53,54]. The extremum of the function $ΔV_Σ$ is determined within the limits of imposed constraints on the heliocentric TOF. In general, the method makes it possible to identify the vicinity of the global minimum of the function under sufficiently loose restrictions on the TOF between the celestial bodies. On the second stage of procedure modification of the Broyden-Fletcher-Goldfarb-Shanno algorithm (BFGS), namely L-BFGS-B algorithm (L is addressed to the limited memory and B is the bound constraint) [55] is used for the precise determination of the function minimum. Note, the algorithm has used the gradient with curvature information for defining direction to a minimum, gradually improving an approximation to the Hessian matrix obtained numerically.

Let us limit the TOF for flight to Sedna, as follows:

$$TOF \leq 20 \text{ yrs} \qquad (1)$$

Emphasise that constraint (1) is acceptable considering that to date, the lifetime of the spacecraft in-orbit is about 15-yr [56,57]. Note that value of constraint (1) is used directly in the optimisation process. Limit on the value of $\Delta V_\Sigma$ will be given further, considering the upper stage's physical (real) characteristics. Note, there would also be used a natural restriction that is the minimum flyby height; by default, this height was taken equal to 5% of the planet radius. Such height was taken because we sought to obtain a maximum gain of the gravity assist keeping the flyby distance in a safe range of any collisions with the planet in the search process.

The application of the described approach to finding optimal trajectories allows to simplify the optimisation process greatly and, at the same time, to obtain good initial approximations for more accurate calculations [49]. The results of [49,60] show that the difference between the minimum $\Delta V$ values obtained using the patched conic approximation and the accurate n-body problem solution is relatively small. However, the launch and flybys dates that correspond to the values may be slightly different [49].

*3.2 Analysis of JGA for flight to Sedna*

Voyages outside the heliosphere are more feasible using the advantages of the gravitational fields of the giant planets. Since Jupiter is the most massive planet, its gravity allows getting the most significant increase in orbital energy of the spacecraft. Several flight schemes to Sedna including assist of this giant planet will be considered below.

An advantage of the Earth-Jupiter-Sedna (EJSed) scheme compared to the schemes that would be described below is that EJSed requires lesser gravity assists, which decreases the additional fuel cost of the trajectory corrections during the passage of celestial bodies. On the other hand, a significant disadvantage of such a flight is that the decreasing of $\Delta V_\Sigma$ is lesser among the other schemes. Fig. 2 shows the $\Delta V_\Sigma$ and the flyby velocity of the spacecraft at Sedna ($V_f$) vs TOF at launch windows from 2025 to 2050.

It is possible to limit $\Delta V_\Sigma$, as follows:

$$\Delta V_\Sigma \leq 12 \text{ km}/s \qquad (2)$$

This restriction is provided by taking the maximum allowable $\Delta V$ of 3.5 km/s for one interplanetary stage (7 km/s for the set of two stages) with a specific impulse of ~308-320 s and by launch vehicle upper stage $\Delta V$ of ~4.5-5 km/s (the specific impulse of about 355 s). For all the manoeuvres, it is supposed to use asymmetrical dimethylhydrazine as fuel and nitrogen (II) tetraoxide as oxidiser because these components provide specific impulse of about 310 s.

Note that the restriction (2) is not practically achievable since there have not yet been missions in which such a $\Delta V_\Sigma$ value has been used. However, for our analysis, this value is important as the filter (i.e. not limited directly into optimisation process) whose purpose is to remove trajectories with a deliberately high $\Delta V_\Sigma$ value and reduce the number of analysed flight schemes.

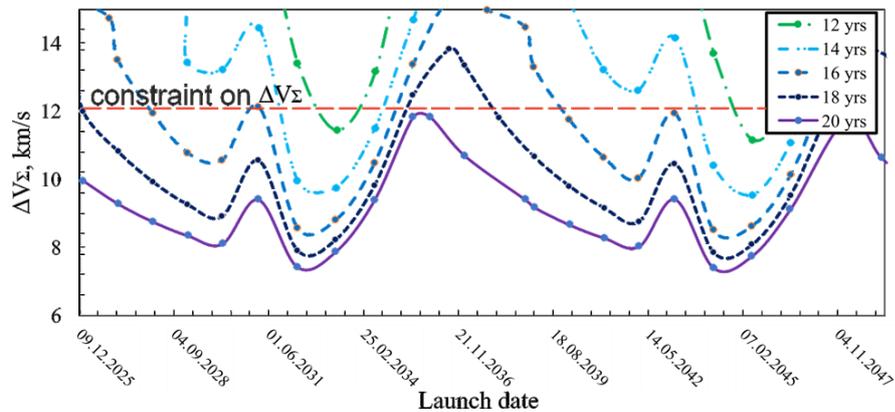

(a)

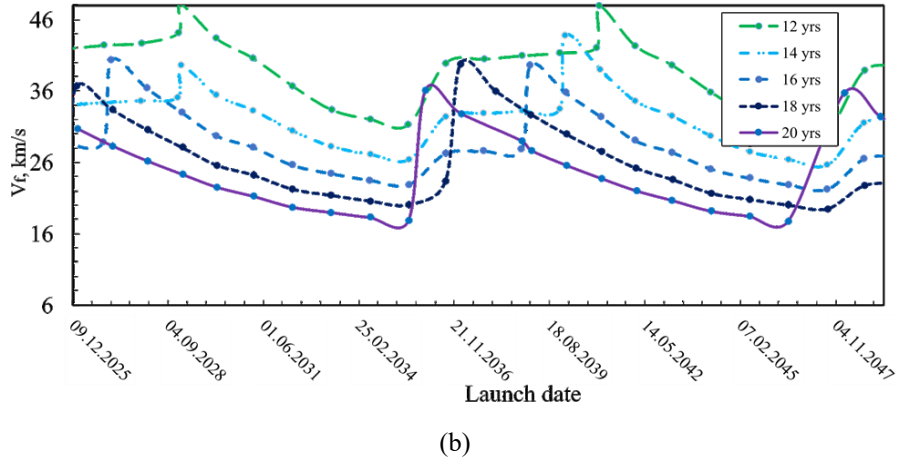

(b)

Fig. 2 (a) $\Delta V_\Sigma$ for 12; 14; 16;18 and 20-yr TOF vs. launch dates, (b) $V_f$ curves for 12; 14; 16;18 and 20-yr TOF vs launch dates.

Looking through the obtained $\Delta V_\Sigma$ profiles in Fig. 2, it is seen that the best launch year is 2032, in which the low $V_f$ value is also obtained (18.9 km/s). The minimal $V_f$ is reached in 2035 and amounts to 17.9 km/s in a 20-yr TOF. If we limit the TOF to 15-yr, the best year changes to 2033 providing the launch $\Delta V$ of 9.3 km/s.

Note the spacecraft flyby Sedna with a high velocity $V_f$ (minimum 17.9 km/sec). Let us give the main reasons for such value of approaching velocity by comparing the mission to Sedna with the New Horizons. In the New Horizons mission [25], the spacecraft passed Pluto at a velocity of 13.7 km/s (distance to Sun of about 30 AU). In our case, however, the speed is 18 km/s or more. If we consider that New Horizons reached Pluto in 9 years at 13.7 km/sec, it takes about 25 years to reach a distance of 80-85 AU. In all the cases considered, reaching Sedna (distance to Sun of about 80 a.u.) is assumed to take 20 years or less. Therefore, the main reason for the high flyby velocity of Sedna is, first of all, a restriction on the flight time. Note, however, that the number of gravity assist maneuvers also has an effect. Since the orbital energy of the spacecraft relative to the Sun will increase (in this case), the approach velocity will also naturally increase, which is confirmed by the data of the tables of Appendix A.1-A.6.

Note that the magnitude of $\Delta V_\Sigma$ is almost entirely repeated every 12 years, i.e. through the orbital period of Jupiter. Such repetition is explained by the fact that Sedna moves only by a small angle on the celestial sphere during a complete revolution of Jupiter and the primary and decisive influence on changings in the dynamics of $\Delta V_\Sigma$ is affected by the mutual position of the Earth and Jupiter, at least during several decades. However, as can be seen from Fig. 2 (a), practically in the vicinity of a $\Delta V_\Sigma$ minimum, namely in 2031 and 2033, there is a rough rise of $\Delta V_\Sigma$. This rough increase is most probably caused by the fact that in these years the angular distance of flight by Earth-Jupiter segment appears to be close to the 180 deg. Because the orbits of the Earth and Jupiter are not coplanar, a considerable increase in the heliocentric orbit inclination (up to more than 7 degrees to the ecliptic) is required. That is possibly the reason for an increase in $\Delta V_\Sigma$ curves in Fig 2.

The detailed catalogue of selected trajectories for different TOF is given in Appendix A, Table A.1. In Fig. 3 the spacecraft's trajectory to Sedna by EJSed scheme at launch in 2032 for 17-yr TOF is shown. The spacecraft trajectory parameters using EJSed scheme for the launch in 2032 are shown in Table 1.

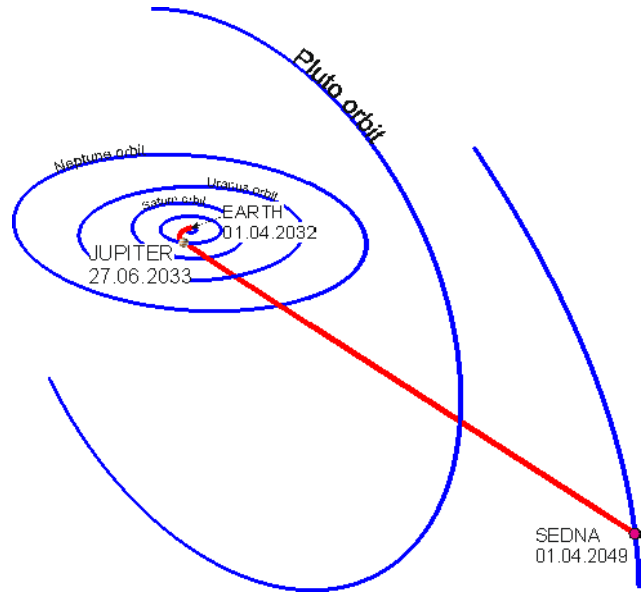

Fig. 3 Trajectory of the spacecraft flight to Sedna by EJSed scheme at launch in 2032

Table 1 The spacecraft trajectory parameters using EJSed scheme for the launch in 2032

| Celestial bodies | Dates of Launch and flyby of celestial bodies, dd.mm.yyyy | Relative velocities near-Earth and of the flyby of celestial bodies, km/s | $\Delta V$ of launch, at aphelion and of the flyby of celestial bodies, km/s | Height of the initial orbit and flyby above celestial bodies, $10^3$ km |
|---|---|---|---|---|
| Earth | 01.04.2032 | 10.93 | 7.73 | 0.2 |
| Jupiter | 27.06.2033 | 16.35 | 0.48 | 4.7 |
| Sedna | 01.04.2049 | 23.77 | - | 0.0* |

\* Approaching an object is assumed to be at any small distance

*3.3 Flight to Sedna with ΔVEGA manoeuvre*

It is possible to use the manoeuvre $\Delta VEGA = \Delta V$ and Earth Gravity Assist, where $\Delta V$ is a small braking impulse in the vicinity of the aphelion on the Earth-Earth loop ($\Delta V_\alpha$) to reduce the cost of the characteristic velocity at the stage of flight to Jupiter. The deceleration at aphelion leads to some increase in the relative velocity during the Earth flyby and enables a significant reduction of $\Delta V$ in the trajectory part Earth-Jupiter.

A study of the flight to Sedna is performed for the Earth-$\Delta V_\alpha$-Earth-Jupiter-Sedna (E$\Delta$VEJSed) scheme compared to the EJSed scheme. Note that the duration of the Earth-Earth trajectory part is determined in the optimisation process by the condition of minimum total $\Delta V$ required for flight to Sedna. In terms of satisfying constraints (1) and (2), the E$\Delta$VEJSed scheme is similar to the EJSed one. Only for launch in 2033, 2034 and 2046, the fulfilment of constraint (2) for E$\Delta$VEJSed scheme is violated (Fig. 4). However, for launch dates 2029 to 2031 and 2035 to 2040, the use of E$\Delta$VEJSed scheme reduces the total $\Delta V$ cost by 1-2 km/s (Fig. 4).

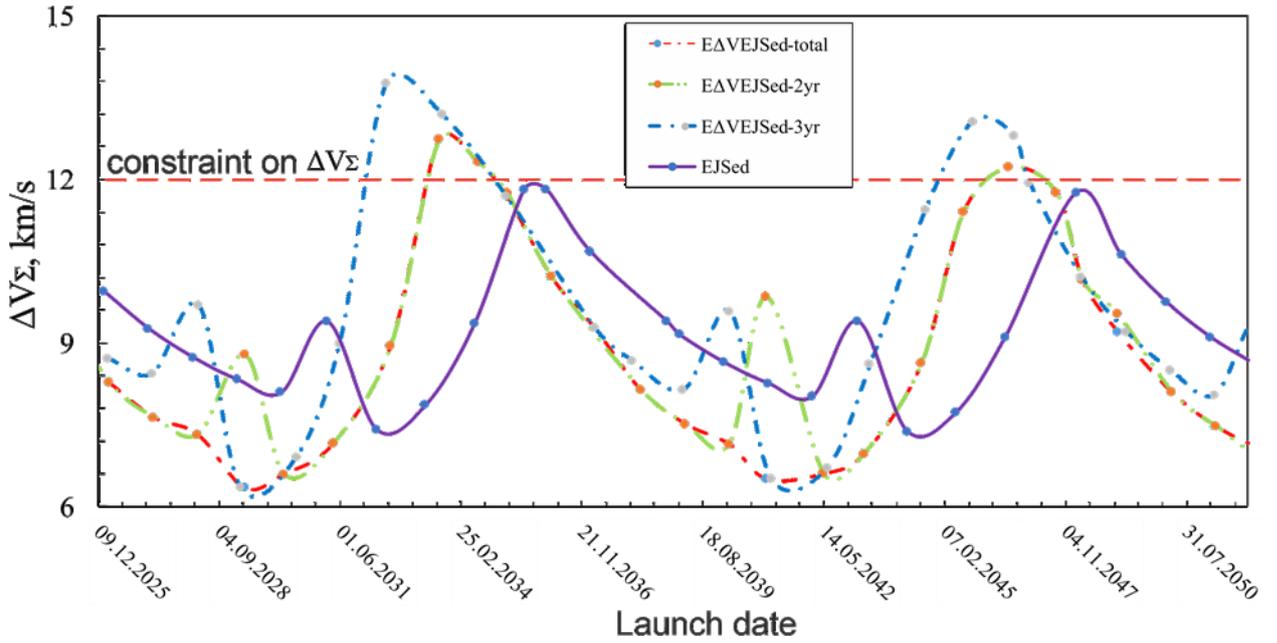

Fig. 4 Comparison of $\Delta V_\Sigma$ vs launch year using EJSed and E$\Delta$VEJSed schemes for 20-yr TOF. In the figure the E$\Delta$VEJSed scheme was separated on the E$\Delta$VEJSed-2yr, which corresponds to the 2-yr flight on Earth-Earth (E-E) section, E$\Delta$VEJSed-3yr which corresponds to the 3-yr on the E-E trajectory part and E$\Delta$VEJSed-total which is consists from the best solutions obtained for 2-yr and 3-yr flight on the E-E trajectory part.

$\Delta V_\Sigma$ vs TOF for the launch dates within 2029-2031 given in Fig. 5. The launch dates from 2029 to 2031 require more detailed consideration since these dates are in the vicinity of $\Delta V_\Sigma$ minimum (Fig. 4). Launch in 2029 allows reducing the $\Delta V_\Sigma$ to ~9 km/s for the 16-yr TOF and to 6.29 km/s for the 20-yr TOF. Such results are comparable with ones obtained in [43,61] for the more complex Earth-Venus-Earth-Earth-Jupiter-Sedna (EVEEJSed) scheme ($\Delta V_\Sigma$ = 6.27 km/s in 2029). Notice that the launch date in 2028 is also in the vicinity of the $\Delta V_\Sigma$ minimum. Still, as it was stated earlier, it makes no practical sense to consider flights before 2029 since the preparation and performance of the mission to Sedna may take more than one decade.

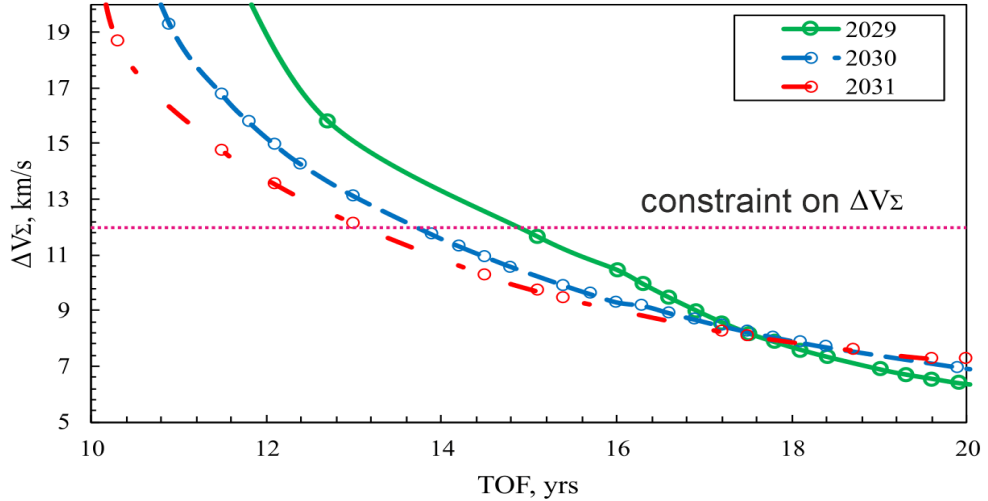

Fig. 5 $\Delta V_\Sigma$ vs TOF for flight to Sedna using E$\Delta$VEJSed scheme at launch in 2029-2031.

The $\Delta V_\Sigma$ curve presented in Fig. 5 shows that the flight to Sedna considering the constraint (2) is achieved for the TOF of 14-14.5 years. A TOF over 16 years will require $\Delta V_\Sigma \approx 9$ km/s similar to the New Horizons [58]. Note that the considered launch years in Fig. 5 TOF of about 17.5 years corresponds to the $\Delta V_\Sigma$ cost is equal to 8.1 km/s.

In Appendix A, Table A.2, the best results for E$\Delta$VEJSed for launch from 2029 to 2042 (one orbital period of Jupiter), satisfying the constraints (1) and (2), are shown. The spacecraft trajectory by the E$\Delta$VEJSed scheme to Sedna in 2029, with a 17-yr TOF, is shown in Fig. 6. The spacecraft trajectory parameters using the E$\Delta$VEJSed scheme at launch in 2029 are shown in Table 2.

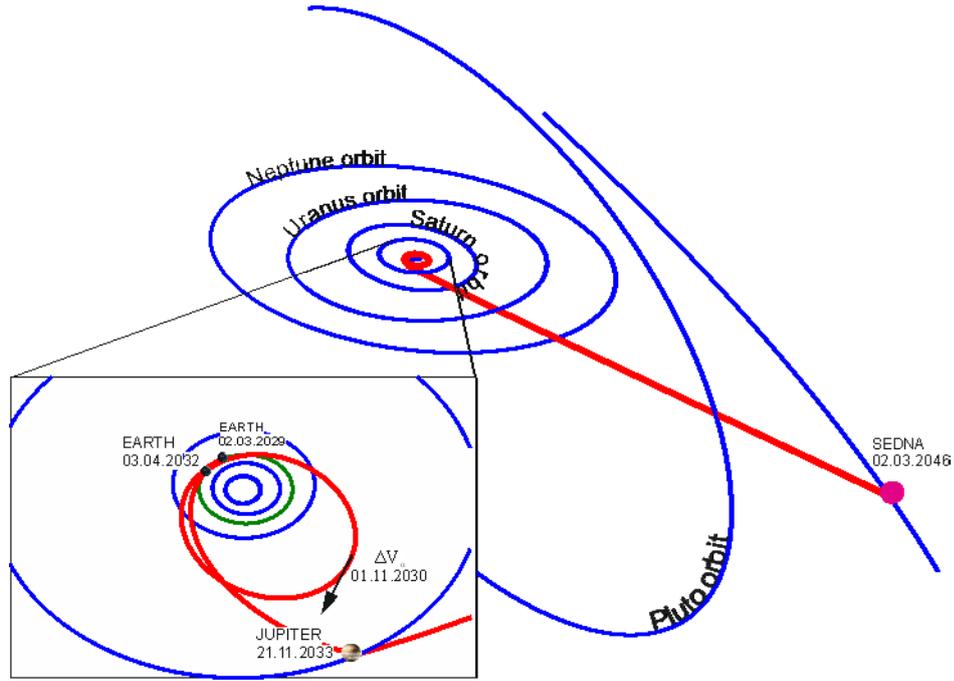

Fig. 6 Spacecraft trajectory to Sedna by EΔVEJSed scheme at launch in 2029

Table 2 The spacecraft trajectory parameters using EΔVEJSed scheme for launch at 2029

| Celestial bodies | Dates of Launch and flyby of celestial bodies | Relative velocities near-Earth and of the flyby of celestial bodies, km/s | ΔV of launch, at aphelion and of the flyby of celestial bodies, km/s | Height of the initial orbit and flyby above celestial bodies, $10^3$ km |
|---|---|---|---|---|
| Earth | 02.03.2029 | 6.94 | 5.24 | 0.2 |
| $\Delta V_\alpha$ | 01.11.2030 | 11.33 | 0.25 | - |
| Earth | 03.04.2032 | 9.54 | 0.00 | 2.7 |
| Jupiter | 21.11.2033 | 21.70 | 3.37 | 3.6 |
| Sedna | 02.03.2046 | 30.53 | - | 0.0* |

\* Approaching an object is assumed to be at any small distance

*3.4 Gravity assist of Neptune in manoeuvre ΔVEGA for the voyage to Sedna*

The expansion of the expedition to Sedna by the additional study of the planets and many of their moons from the flyby trajectory is of particular interest. Of course, such an extension of the mission should not increase TOF and ΔV costs. Studies considering the expansion of gravity assist scenarios have already been made in [23,43]. It has been shown that due to the orbital positions of Uranus and Sedna, the gravity assist of Uranus is entirely impossible in the next few decades. On the other hand, using Saturn is possible only at the flight duration of 33-yr or more [43]. However, in [43] in the number of schemes Neptune flyby allows reducing the value of $\Delta V_\Sigma$ significantly, and in some cases allows increasing the altitude of the Jupiter flyby, what is also important.

The current paper considers the use of the Earth-Jupiter-Neptune-Sedna (EJNSed) and Earth-ΔV-Earth-Jupiter-Neptune-Sedna (EΔVEJNSed) schemes. The first scheme slightly improves the EJSed scheme, while the second scheme combines the advantages of ΔVEGA, JGA, and Neptune's gravity assist.

Let us first consider the EΔVEJNSed scheme. Fig. 7 shows the $\Delta V_\Sigma$ vs launch date for 20-yr and 30-yr TOF using the EΔVEJNSed scheme and only for 20-yr TOF using the EJSed scheme. The results for the EJSed scheme are given to compare both schemes by the $\Delta V_\Sigma$ value.

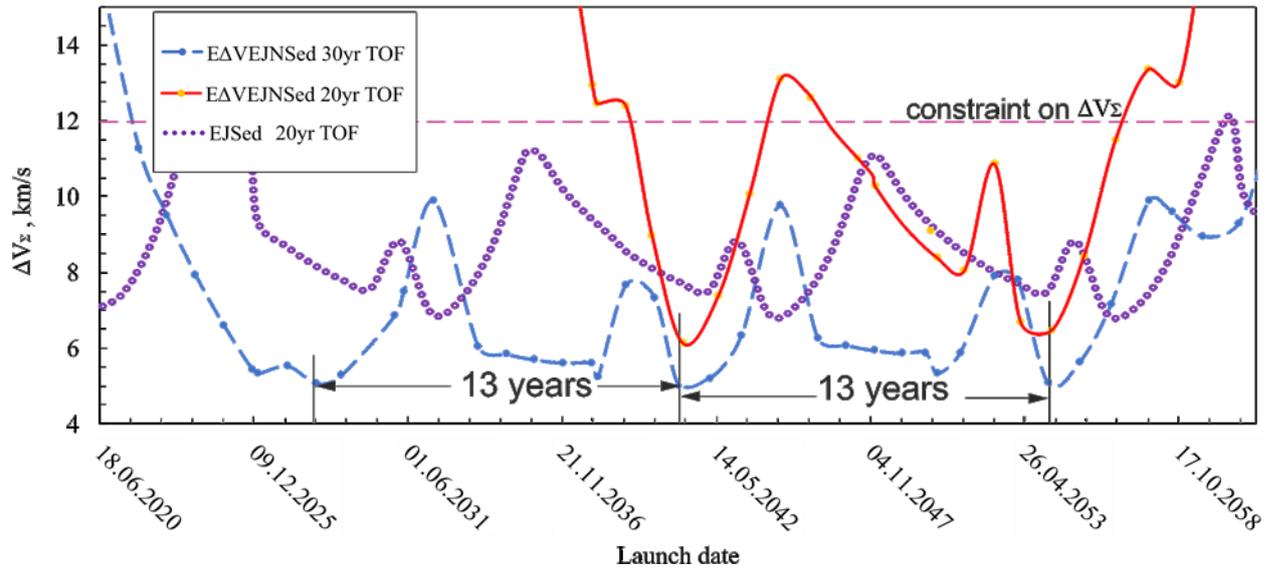

Fig. 7 $\Delta V_\Sigma$ vs launch date for E$\Delta$VEJNSed scheme for TOF of 20 and 30 yrs

In the considered launch date interval from 2029 to 2040, the use of the E$\Delta$VEJNSed scheme under the constraint (1) does not give a significant gain compared to the EJSed and E$\Delta$VEJSed schemes. Still, in 2041, the E$\Delta$VEJNSed scheme may be the best for 20-yr TOF among other considered schemes, including ones in the papers [43]. Such result cannot be due only to a random coincidence of the lucky orbital positions of the Earth, Jupiter, Neptune and Sedna, i.e. there must be a regularity of the appearance of such minima. And such explanation has been found. In the research of Alexander A. Sukhanov,[2] the calculation of the repeatable configuration for any number of planets was considered. For the combination of Earth, Jupiter and Neptune a period of repeating their mutual positions of 1, 12 and 13 yrs was obtained, depending on the required accuracy of the calculation. It is easy to see that a flight using gravity assists of these planets is possible every year following the abovementioned results. However, in this problem, the influence of the position of Sedna relative to Jupiter and Neptune seems to be the decisive factor, so a low-cost flight is not possible every year. However, the recurrence frequency of minimum cost ($\Delta V_\Sigma$ = 6.13 km/s in 2041) is achievable every 13 years. A confirmation of this result is the flight in 2054 (Fig. 7), the minimum cost $\Delta V_\Sigma$ for which will be about 6.48 km/s. The flight in 2028 can also be performed with costs comparable to those of 2041 but with a TOF of at least 30 years (Fig. 7).

A catalogue of trajectories using E$\Delta$VEJNSed scheme and satisfying (1) and (2) is given in Appendix A, Table A.3. The spacecraft trajectory by E$\Delta$VEJNSed scheme to Sedna in 2041, with an 18-yr TOF, is shown in Fig. 8. The spacecraft trajectory parameters using the E$\Delta$VEJNSed scheme at launch in 2041 are shown in Table 3.

---

[2] His paper "Repeatability of a given configuration of planets" is in preparation.

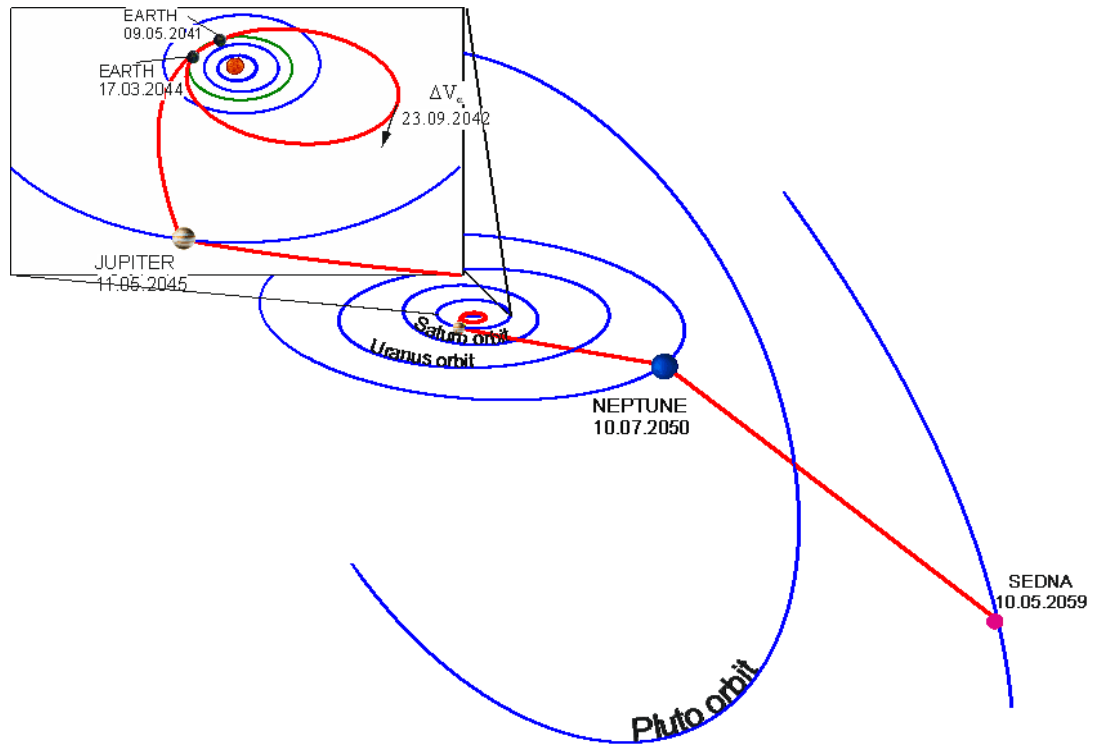

Fig. 8 Trajectory of the spacecraft flight to Sedna in the EΔVEJNSed scheme at launch in 2041.

Table 3. The spacecraft trajectory parameters using EΔVEJNSed with the TOF of 18-yr

| Celestial bodies | Dates of Launch and flyby of celestial bodies | Relative velocities near-Earth and of the flyby of celestial bodies, km/s | ΔV of launch, at aphelion and of the flyby of celestial bodies, km/s | Height of the initial orbit and flyby above celestial bodies, $10^3$ km |
|---|---|---|---|---|
| Earth | 09.05.2041 | 7.05 | 5.30 | 0.2 |
| $\Delta V_\alpha$ | 23.09.2042 | 11.14 | 0.57 | - |
| Earth | 17.03.2044 | 13.31 | 1.01 | 0.3 |
| Jupiter | 11.05.2045 | 17.79 | 0.32 | 42.6 |
| Neptune | 10.07.2050 | 25.59 | 1.81 | 1.2 |
| Sedna | 10.05.2059 | 27.90 | - | 0.0* |

* Approach to the object is assumed to be at any small distance

*3.5 Gravity assist of Neptune in Earth-Jupiter-Sedna voyage*

It is almost impossible to perform a flight using EJNSed for launch dates up to 2039 under the constraints (1) and (2), as shown in Fig. 9. However, for launch dates from 2040 to 2060, a few launch windows ensuring minimum $\Delta V_\Sigma$ satisfying constraint (2) are found (Fig. 9).

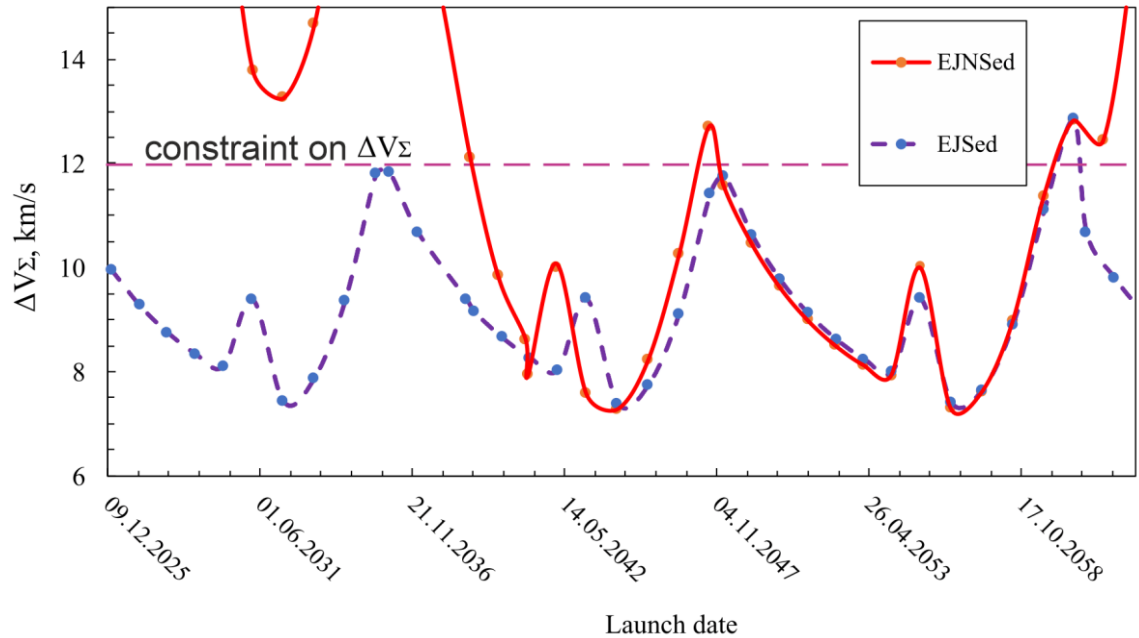

Fig. 9 Comparison of $\Delta V_\Sigma$ for 20-yr of TOF vs launch dates for EJSed and EJNSed schemes.

The EJNSed flight for 20-yr TOF shows no significant advantages comparing with the EJSed scheme (Fig. 9). Using the EJNSed scheme at launch in 2041 allows reducing the $\Delta V_\Sigma$ value on 330 m/s compared with the EJSed scheme. The period of repeatability of the $\Delta V_\Sigma$ is equal to 13 years and is the same as for the previously discussed E$\Delta$VEJNSed scheme.

Table A.4 (Appendix A) contains a catalogue of selected trajectories to Sedna using the EJNSed scheme. In Fig. 10, the trajectory to Sedna using the EJNSed scheme at launch in 2044, with a duration of 20-yr is shown. The trajectory characteristics of the flight to Sedna using the EJNSed scheme are shown in Table 4.

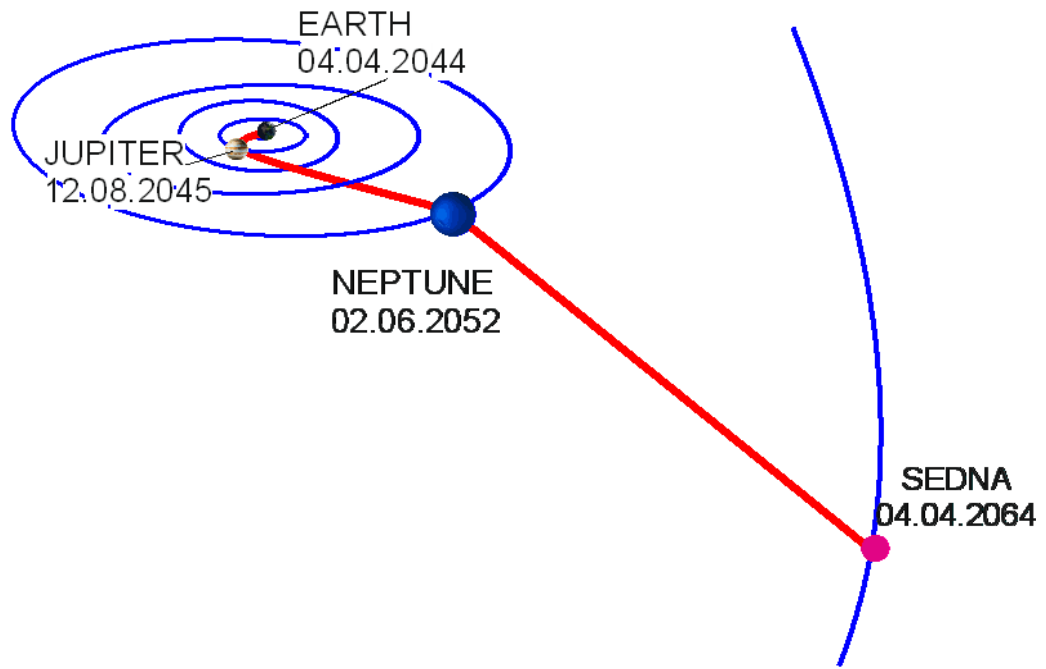

Fig. 10 Spacecraft trajectory to Sedna using the EJNSed scheme for launch in 2044

Table 4. The spacecraft trajectory parameters using EJNSed scheme for launch in 2044 with the 20-yr TOF

| Celestial bodies | Dates of Launch and flyby of celestial bodies, dd.mm.yyyy | Relative velocities near-Earth and of the flyby of celestial bodies, km/s | ΔV of launch, at aphelion and of the flyby of celestial bodies, km/s | Height of the initial orbit and flyby above celestial bodies, $10^3$ km |
|---|---|---|---|---|
| Earth | 04.04.2044 | 10.29 | 7.29 | 0.2 |
| Jupiter | 12.08.2045 | 12.58 | 0.00 | 232.4 |
| Neptune | 02.06.2052 | 18.43 | 0.00 | 20.0 |
| Sedna | 04.04.2064 | 20.34 | - | 0.0* |

\* Approach to the object is assumed to be at any small distance

*3.6 Remark on the VEEGA and VEΔVEGA manoeuvres in the problem of the fastest approach to Sedna*

In the paper [43], the authors considered the use of different flight schemes to Sedna based on the VEEGA (Venus-Earth-Earth Gravity Assist) manoeuvre or its modification VEΔVEGA (modification of VEEGA with breaking ΔV in the aphelion of Earth-Earth loop) at the launch dates in 2029-2034. An advantage of these schemes is that in some cases the flight would only require ΔV necessary for the flight to Venus. However, the TOF interval of 20 to 50 yrs adopted by the authors [43] does not fully evaluate the effectiveness of this approach in finding the fastest route to reach Sedna under the constraints (1). For this reason, in the current paper the $\Delta V_\Sigma$ vs TOF was plotted for the launch in 2029 only for the interval of TOF from 10 to 20 yrs (Fig. 11).

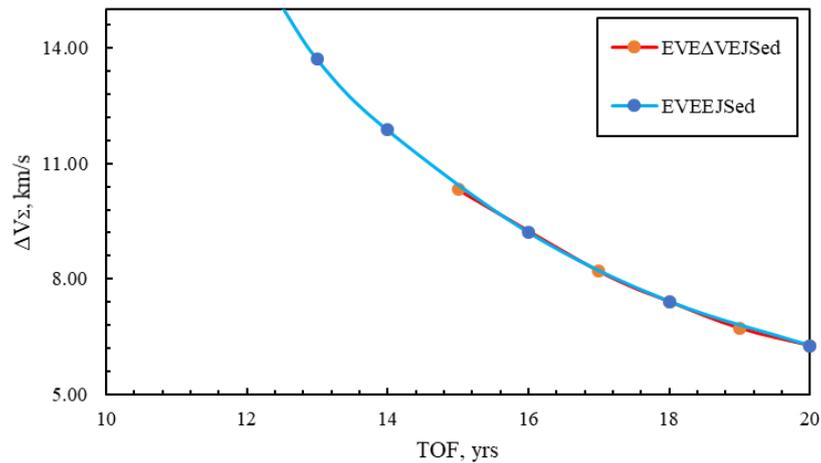

Fig. 11 $\Delta V_\Sigma$ vs TOF for flight to Sedna using EVEEJSed and EVEΔVEJSed schemes at launch in 2029

Flight to Sedna using the EVEEJSed and EVEΔVEJSed schemes under the constraints (1) and (2) is significantly better in terms of $\Delta V_\Sigma$ than EJSed scheme beginning from 14 yrs TOF (Fig. 11). However, it should be emphasised that EVEEJSed and EVEΔVEJSed are significantly more complex schemes than any other ones described in this paper, which will eventually lead to higher costs for correction manoeuvres for flight using these schemes.

The catalogue of selected trajectories to Sedna using EVEEJSed and EVEΔVEJSed schemes satisfying constraints (1) and (2) is given in Appendix A, Table A.5. Fig. 12 shows the trajectory to Sedna using EVEΔVEJSed scheme for the TOF of 17-yr (Table A.5). The spacecraft trajectory parameters using the EVEΔVEJSed scheme are shown in Table 5.

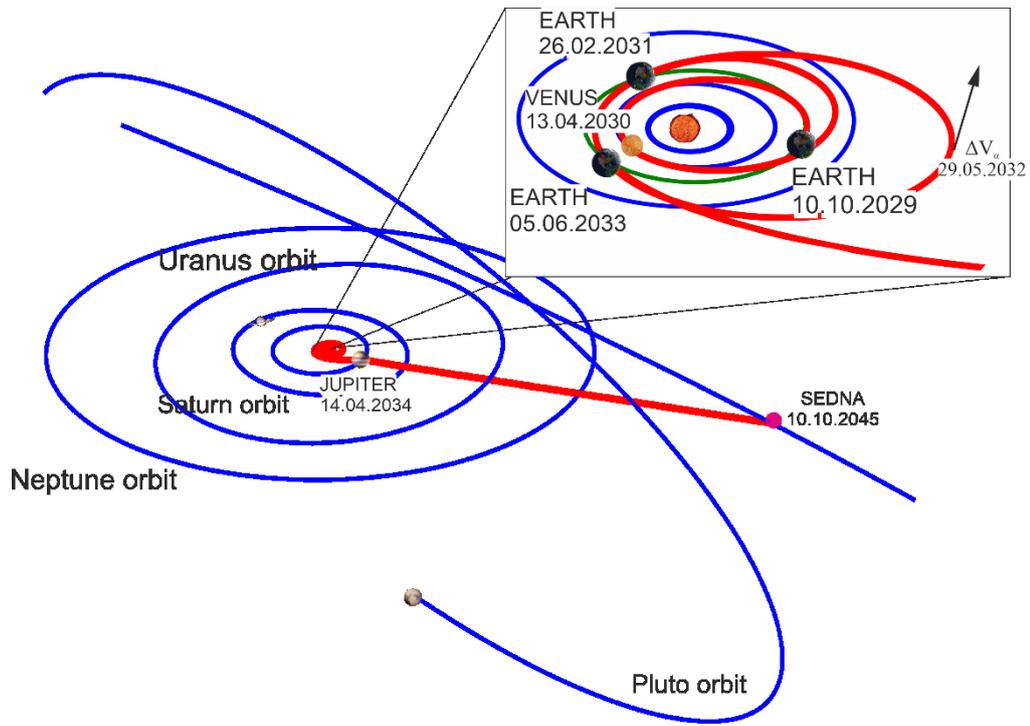

Fig. 12 Trajectory to Sedna using the EVEΔVEJSed scheme at launch in 2029

Table 5. The spacecraft trajectory parameters using EVEΔVEJSed scheme for launch at 2029 with the 17-yr TOF

| Celestial bodies | Dates of Launch and flyby of celestial bodies, dd.mm.yyyy | Relative velocities near-Earth and of the flyby of celestial bodies, km/s | ΔV of launch, at aphelion and of the flyby of celestial bodies, km/s | Height of the initial orbit and flyby above celestial bodies, $10^3$ km |
|---|---|---|---|---|
| Earth | 10.10.2029 | 3.56 | 3.80 | 0.2 |
| Venus | 13.04.2030 | 5.62 | 0.00 | 7.7 |
| Earth | 26.02.2031 | 9.73 | 0.00 | 5.7 |
| $\Delta V_\alpha$ | 29.05.2032 | 14.74 | 0.00 | - |
| Earth | 05.06.2033 | 15.80 | 4.54 | 0.3 |
| Jupiter | 14.04.2034 | 23.66 | 0.87 | 52.1 |
| Sedna | 10.10.2045 | 32.27 | - | 0.0* |

* Approach to the object is assumed to be at any small distance

4. **A flight to Sedna with the Oberth manoeuvre near the Sun**

In this paper, the following scheme of the flight is considered: the spacecraft, with the help of planet gravity assists approaches the Sun at the perihelion distance from 2 to 10 Solar radii and then performs a manoeuvre required to make the value of the heliocentric velocity of the spacecraft sufficient for the flight to Sedna under the constraints (1) and (2). The optimisation problem is again to find the minimum of $\Delta V_\Sigma$.

However, the main disadvantage of such scheme compared with other considered in this paper schemes is worth pointing out here. The close approach the spacecraft to the Sun would demand high-level thermal protection for spacecraft. Otherwise, a well-known example of the Parker Solar Probe mission (NASA), approaching the Sun, is supposed to be at ten solar radii (~7 million km) in April 2024. Despite the low pericentre height, Parker's thermal shield is only ~13% of its total mass[3], and the shield design can significantly reduce the spacecraft's heating [31,59,60].

---

[3] 72.9 kg out of total mass of the spacecraft ~555 kg [59]

In Fig. 13, a scheme of the flight to Sedna using OM is shown. The flight to Jupiter takes place at first. Then the spacecraft trajectory is reversed towards the Sun, with a subsequent OM in the vicinity of perihelion and then the spacecraft flights to Sedna.

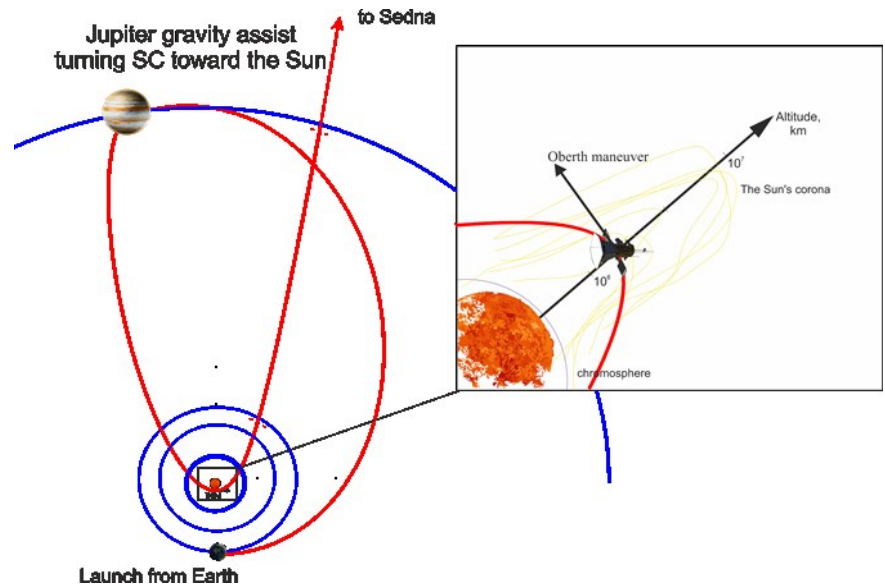

Fig. 13 Scheme of flight to Sedna with an OM near the Sun

The $\Delta V_\Sigma$ and $V_f$ vs TOF obtained using the Earth-Jupiter-OM-Sedna (EJ-OM-Sed) scheme are plotted in Fig. 14 for a few decades. This period was chosen for the purpose of comparison EJ-OM-Sed and EJSed schemes.

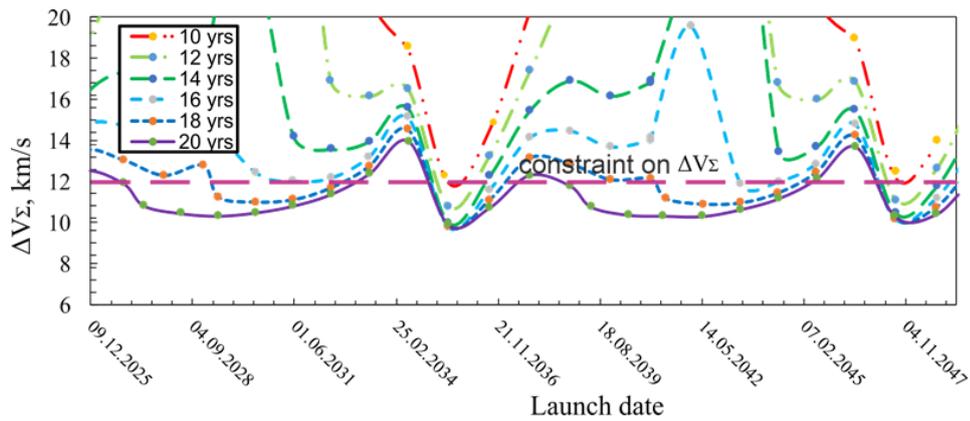

(a)

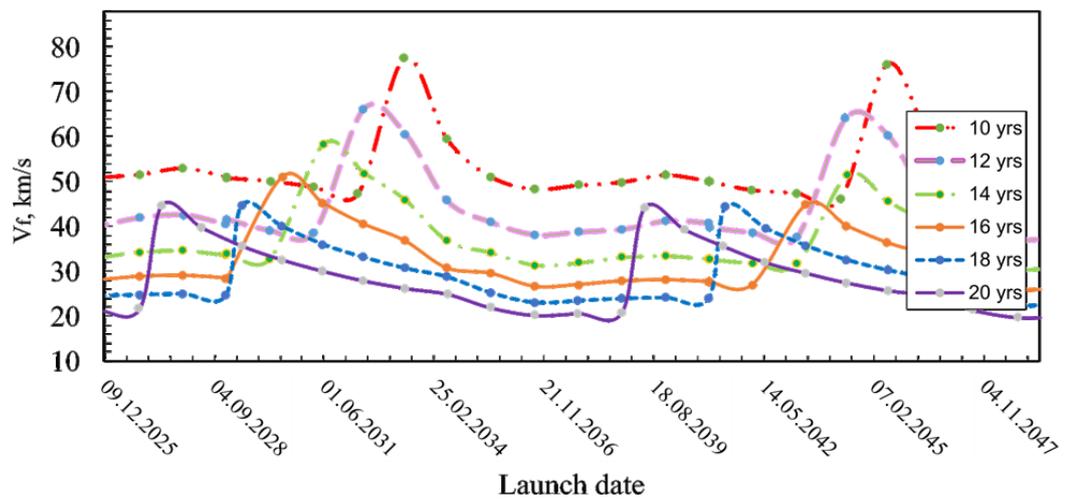

(b)

Fig. 14 (a) ΔV_Σ for 10; 12; 14; 16;18 and 20-yr TOF vs. launch dates, (b) $V_f$ curves for 10; 12; 14; 16;18 and 20-yr TOF vs launch dates.

Notice, the flight using the EJ-OM-Sed scheme (Fig. 14.a) is almost opposite to the one using the EJSed scheme (Fig. 2.a). From the other hand the flight using EJ-OM-Sed requires higher ΔV_Σ than using EJSed scheme. Let us compare the best launch dates for the flight using EJSed in 2032 (Fig 2.a) and EJ-OM-Sed in 2035 (Fig. 14.a).

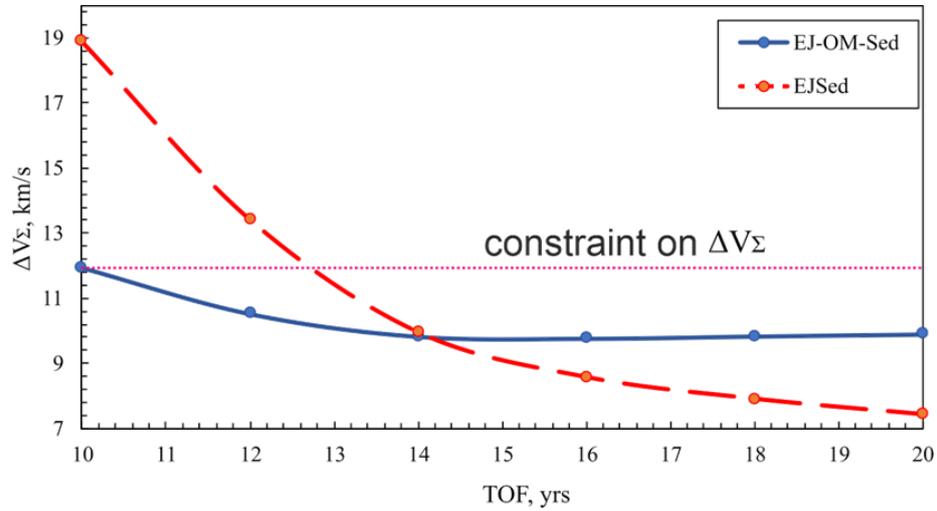

Fig. 15 Comparison of EJSed and EJ-OM-Sed schemes by ΔV_Σ for the best launch dates in 2032 and 2035, respectively

The application of such a scheme with OM (i.e. EJ-OM-Sed in Fig. 15) gives the greatest benefit for a TOF less than 14-yr (Table 6) compared with the other mentioned in this paper schemes. However, since ΔV_Σ using the scheme is always greater than 9.8 km/s for any launch date out of the period under consideration, the scheme does not seem appropriate for the TOF of 15-yr or more.

As it was mentioned above, the requirement of ensuring the extra thermal shield of the spacecraft reduces its final mass by about 15% that is the main disadvantage of the used scheme. However, it should be emphasised that the flight was considered only with the high thrust, but the situation might completely change in the case of using the low thrust. For the low thrust, such proximity (Table 6) to the Sun may be an advantage, as it allows the use of promising technology, namely a solar sail. As it was shown in studies [30,31,39,42,61], the application of such technology will enable the spacecraft to accelerate up to 20-25 AU/y (94.8-118.5 km/s) manoeuvring near the Sun, while the best result obtained with the high thrust is only 8-11 AU/y (37.9-52.1 km/s) at a very close approach to the Sun and the ΔV manoeuvre value of ~3-4 km/s. Increasing the final velocity to 20-25 AU/y would require an impulse of at least 8 km/s, which is impractical for our target.

Table 6. Catalogue of selected trajectories to Sedna by EJ-OM-Sed scheme satisfying constraints (1) and (2)

| Optimal launch date, dd.mm.yyyy | TOF, yrs | Jupiter flyby date, dd.mm.yyyy | Height of Jupiter flyby, $10^3$ km | ΔV_Σ, km/s | $V_f$, km/s | $r_{sun,*}$ $10^6$ km |
|---|---|---|---|---|---|---|
| 15.05.2029 | 18 | 07.02.2037 | 951.1 | 11.24 | 44.83 | 2.3 |
| 11.05.2029 | 20 | 28.12.2036 | 859.2 | 10.32 | 35.75 | 2.4 |
| 16.05.2030 | 18 | 14.01.2037 | 973.7 | 11.00 | 40.04 | 2.3 |
| 15.05.2030 | 20 | 09.01.2037 | 966.5 | 10.49 | 32.64 | 2.6 |
| 22.05.2031 | 16 | 06.02.2037 | 1,220.0 | 12.00 | 45.31 | 2.3 |
| 18.05.2031 | 18 | 24.12.2036 | 1,089.3 | 11.10 | 36.12 | 2.5 |
| 19.05.2031 | 20 | 10.01.2037 | 1,109.9 | 10.80 | 30.04 | 3.5 |
| 24.05.2032 | 16 | 16.01.2037 | 1,521.7 | 12.15 | 40.52 | 2.3 |
| 24.05.2032 | 18 | 19.01.2037 | 1,480.9 | 11.67 | 33.10 | 3.1 |
| 26.05.2032 | 20 | 07.02.2037 | 1,446.2 | 11.37 | 27.97 | 4.7 |
| 24.07.2035 | 10 | 06.10.2036 | 178.4 | 11.95 | 51.35 | 2.3 |
| 23.07.2035 | 12 | 20.10.2036 | 308.8 | 10.53 | 41.21 | 2.3 |
| 23.07.2035 | 14 | 30.10.2036 | 437.9 | 9.83 | 34.22 | 2.3 |
| 22.07.2035 | 16 | 11.11.2036 | 390.0 | 9.77 | 29.53 | 2.4 |

| 22.07.2035 | 17 | 30.11.2036 | 404.4 | 9.84  | 25.23 | 3.1  |
| 21.07.2035 | 19 | 14.12.2036 | 423.9 | 9.90  | 21.96 | 3.7  |
| 29.08.2036 | 14 | 17.10.2037 | 3.7   | 12.25 | 31.41 | 11.1 |
| 29.08.2036 | 16 | 18.10.2037 | 3.6   | 11.58 | 26.59 | 13.0 |
| 29.08.2036 | 18 | 23.10.2037 | 7.3   | 11.09 | 23.02 | 15.1 |
| 29.08.2036 | 20 | 31.10.2037 | 16.3  | 10.72 | 20.27 | 17.1 |
| 28.10.2038 | 20 | 24.11.2040 | 3.6   | 11.78 | 20.94 | 2.3  |
| 28.11.2039 | 20 | 28.01.2042 | 3.6   | 11.01 | 21.13 | 2.3  |
| 24.05.2040 | 18 | 18.04.2049 | 975.2 | 11.81 | 50.35 | 2.3  |
| 20.05.2040 | 20 | 17.03.2049 | 839.7 | 10.36 | 39.30 | 2.3  |

*perihelion distance of the spacecraft

Table 7 shows the spacecraft trajectory parameters of the flight to Sedna, using the EJ-OM-Sed scheme, at launch in 2035. Fig. 16 shows the trajectory of the corresponding flight.

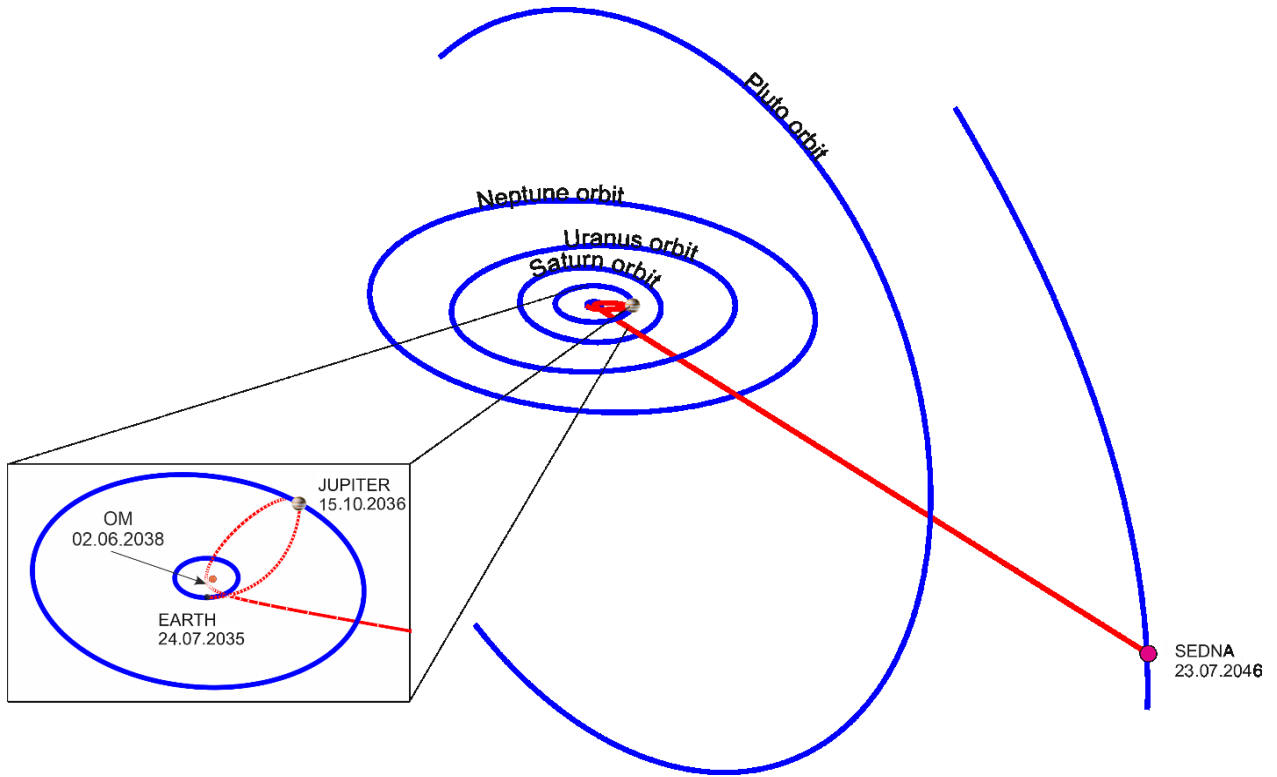

Fig. 16 Trajectory of spacecraft flight to Sedna for the launch in 2032 using EJ-OM-Sed scheme

Table 7. The spacecraft trajectory parameters using EJ-OM-Sed scheme with 11-yr TOF for launch in 2035

| Celestial bodies | Dates of Launch and flyby of celestial bodies, dd.mm.yyyy | Relative velocities near-Earth and of the flyby of celestial bodies, km/s | ΔV of launch, at aphelion and of the flyby of celestial bodies, km/s | Height of the initial orbit and flyby above celestial bodies, $10^3$ km |
|---|---|---|---|---|
| Earth    | 24.07.2035 | 11.24  | 7.94 | 0.2     |
| Jupiter  | 15.10.2036 | 13.93  | 0    | 244.5   |
| OM (Sun) | 02.06.2038 | 367.00 | 3.17 | 2,000.0 |
| Sedna    | 23.07.2046 | 45.79  | -    | 0.0*    |

* Approach to the object is assumed to be at any small distance

As shown in Table 7, while approaching the Sun the spacecraft accelerates to the 367 km/s, and while approaching Sedna its velocity $V_f$ is about 45.79 km/s which is higher than in all other schemes (Table 1-5). Considering this value, the flyby Sedna would be fast, resulting in blurring images taken by on-board cameras.

5. **Estimation of the spacecraft's mass delivered to Sedna**

The final mass of the spacecraft delivered to Sedna was estimated using the rocket equation [62]. For this estimation, we used the launch vehicles presented in Table 8.

Table 8. Characteristics of analysed launch vehicles

| Rocket | Mass in the Low Earth Orbit, $10^3$ kg | Fuel capacity, $10^3$ kg | Dry mass of the last stage, $10^3$ kg | The specific impulse of the last stage, s |
|---|---|---|---|---|
| Soyuz 2.1.b / Fregat [*] (fourth stage) | 8.2 | ~5.6 | 0.97 | 333.2 |
| Proton-M / DM-03[**] (fourth stage) | 23.7 | ~18.0 | 2.34 | 355 |
| Delta IV Heavy[***, 1]/DCSS (second stage) | 28.8 | 27.2 | 3.48 | 462 |
| Space Launch System (SLS)[***, 1]/ ICPS (second stage) | 95.0 | 27.2 | 3.50 | 465.5 |

[*] - performance is given under the condition of launch from Vostochny site
[**] - performance is given under the condition of launch from Baikonur site
[****] - performance is given under the condition of launch from Cape Canaveral (airforce base) site
[1] – performance curves obtained in works [63] and through launch calculator by https://elvperf.ksc.nasa.gov/Pages/Default.aspx

It should be noted that it is assumed that a chemical upper stage is used for performing manoeuvres on the interplanetary arcs in all cases. The analysed upper stage has a specific impulse of 308 s, which is provided by using two components fuel: asymmetrical dimethylhydrazine (fuel) and nitrogen tetroxide (oxidiser). The constant mass of the upper stage is 110 kg. The mass of the tanks is calculated in 15% out of the fuel mass.

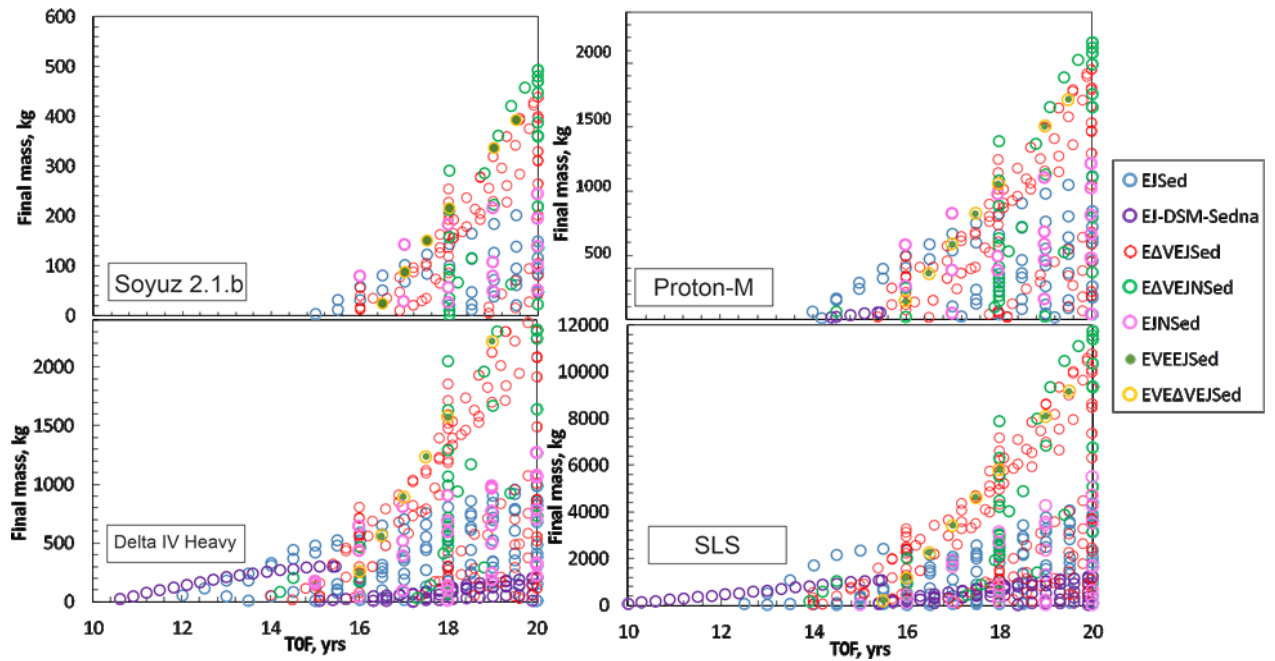

Fig. 17 The final mass brought to Sedna vs TOF for every considered scheme and considered launch vehicles

The distribution of the final mass in Fig. 17, depending on the used scheme and the used launch vehicle, demonstrates the technical possibility of a flight to Sedna under conditions (1) and (2). Using a middle-class launch vehicle (Soyuz 2.1.b), the payload mass delivered to Sedna would be up to 550 kg for a 20-year TOF. This value is twice more than for Pioneer 10, Pioneer 11 and is a half of the mass of Voyager 1, Voyager 2. However, using the super-heavy class, it is possible to deliver to Sedna up to 12,000 kg of the payload, which makes an orbiting Sedna theoretically possible. However, due to the high flyby velocity of the spacecraft, this payload mass still might not be enough to produce a braking impulse comparable to the flyby velocity (>20 km/s).

6. **Remark on the close Jupiter flyby**

Almost all the schemes used in the research require the close flyby of Jupiter at a distance of less than 1 million km. In such a case, the spacecraft might be exposed due to an intense radiation load. The authors of the research [64], based on the dipole approximation of the magnetic field model of Jupiter ('O4' model by [65]), estimated the vehicle radiation exposure during the Jupiter flyby. According to the results obtained, a Jupiter flyby without causing damage of an on-board equipment is possible if the flight altitude and inclination are greater than 136 thousand km and 40 degrees [64], respectively. Later studies [66] show that a safe passage at any inclination of the spacecraft orbit is possible at an altitude over 600 thousand km (8.4 radii of Jupiter).

The analysis prepared in this paper was conducted using free SPENVIS software[4] for the Divine & Garret radiation model [67] at an average trajectory inclination of 25 degrees to the Jupiter equatorial plane. The simulation results for the aluminium shielding are shown in Fig 18. It is shown that a Jupiter flyby is possible only if the permanent radiation shielding of the spacecraft is increased by at least 10-20 mm. In that case, radiation will be about 100-200 krad which is close to the Galileo exposure ~290 krad. Note that this thickening of the shielding leads to an increase in mission cost and a reduction in payload mass.

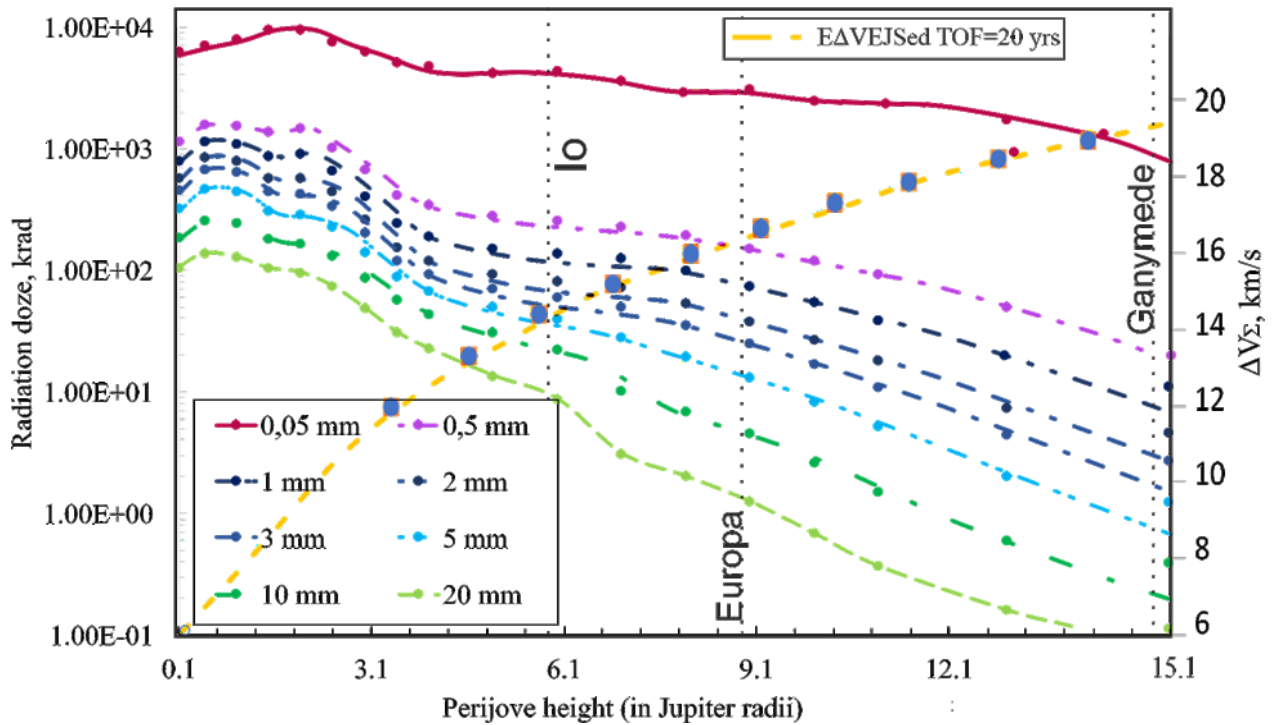

Fig. 18 Total radiation dose and $\Delta V_\Sigma$ cost vs perijove height

The height of the Jupiter flyby can be raised to any desired value, but this inevitably would lead to increasing $\Delta V$ costs. For example, Fig. 18 shows an estimate for the increase in total $\Delta V\Sigma$ vs perijove height during a flight according to the E$\Delta$VEJSed scheme. This Fig. 18 shows that the increasing flyby pericenter altitude to the height of Ganymede orbit would be accompanied with an over 3-fold increase in $\Delta V_\Sigma$. Taking this into account, the best solution, in this case, would be to increase the shielding thickness of the spacecraft and flyby at a low altitude near Jupiter than to increase the flyby altitude of the spacecraft.

7. **The best flight schemes to Sedna**

Above in this paper, schemes of flight to Sedna fulfilling restrictions (2) and (3) were considered. However, it should be noted that a flight under constraint (3) represents a relatively complicated technical task or may require the use of heavy or super-heavy launch vehicle. Therefore, it is reasonable to select a less energy expansive scheme, i.e. to decrease the constraint (3) down to ~9 km/s that is similar to the $\Delta V$ used in New Horizons[5] mission [25,27,58]. Table 9 shows a catalogue of such schemes. Mass estimates are given only for medium and heavy launch vehicles (Soyuz 2.1.b, Proton-M) since the $\Delta V_\Sigma$ limit allows the use of less expensive launch vehicles than super-heavy ones (such as Space Launch System (SLS)).

---

[4] SPENVIS software web page: https://www.spenvis.oma.be/models.php (Accessed 20 September, 2021)
[5] New Horizons mission web-page URL: https://www.nasa.gov/mission_pages/newhorizons/main/index.htmls

Table 9. Catalogue of the best schemes of flight to Sedna based on ΔV$_\Sigma$

| Scheme | Launch date, dd.mm.yyyy | TOF, yrs | ΔV$_\Sigma$, km/s | Final mass delivered to Sedna | | V$_f$, km/s |
|---|---|---|---|---|---|---|
| | | | | Soyuz 2.1.b, 10$^3$ kg | Proton-M, 10$^3$ kg | |
| EJSed | 31.03.2032 | 20 | 7.42 | 0.22 | 1.07 | 19.73 |
| EJNSed | 04.04.2044 | 20 | 7.29 | 0.24 | 1.15 | 20.34 |
| EΔVEJSed | 02.03.2029 | 20 | 6.36 | 0.44 | 1.87 | 23.91 |
| EΔVEJNSed | 08.03.2041 | 20 | 6.13 | 0.49 | 2.07 | 24.59 |
| EVEEJSed & EVEΔVEJSed | 08.11.2029 | 20 | 6.27 | 0.47 | 1.90 | 23.60 |

The scheme with OM near the Sun does not satisfy constraint by 9 km/s but provides a flight to Sedna with 10-14 yrs duration satisfying the constraint (2). That is the primary reason why this scheme is considered in Table 10. Note, for the EJ-OM-Sed scheme in Table 10, the mass would be considered only for heavy and super-heavy launch vehicles (Delta IV Heavy, Space Launch System (SLS)). Results for other schemes are given in Table 10 considering the ~9 km/s limits.

Table 10. The fastest routes to Sedna

| Scheme | Launch date, dd.mm.yyyy | TOF, yrs | ΔV$_\Sigma$, km/s | Final mass delivered to Sedna, 10$^3$ kg | V$_f$, km/s |
|---|---|---|---|---|---|
| EJSed | 29.03.2032 | 15.1 | 9.02 | 0.30 (Proton-M) | 27.49 |
| EJNSed | 10.04.2044 | 15.7 | 9.07 | 0.29 (Proton-M) | 26.67 |
| EΔVEJSed | 24.02.2042 | 16.0 | 8.95 | 0.31 (Proton-M) | 29.96 |
| EΔVEJNSed | 09.05.2041 | 18.0 | 8.99 | 0.31 (Proton-M) | 27.90 |
| EVEEJSed & EVEΔVEJSed | 02.11.2029 | 16.2 | 9.05 | 0.22 (Proton-M) | 31.60 |
| EJ-OM-Sed | 24.07.2035 | 10.0 | 11.95 | 0.10 (SLS) | 51.35 |

The best scheme providing the smallest value of ΔV$_\Sigma$ under constraint (2) is the EΔVEJNSed scheme enabling ΔV$_\Sigma$ =6.13 km/s for the TOF of 20-yr at launch in 2041 (Table 9). Note that due to the gravity assist of Neptune, there are no earlier launch windows ensuring ΔV$_\Sigma$ =6.13 km/s than in 2041. The next available launch window will be in 2054, after which the ΔV$_\Sigma$ for TOF of 20-yr will steadily increase. Therefore, for a launch before 2041, the EVEΔVEJSed (EVEEJSed) scheme for a launch date in 2029 appears most appropriate, since in that case, a 20-year flight would require ΔV$_\Sigma$ =6.27 km/s (Table 9). The trajectories of the spacecraft flight using EΔVEJNSed and EVEΔVEJSed schemes are given in Fig 8 and Fig 12.

If reaching Sedna in the shortest time while satisfying constraint (1) is needed, it is better to use the EJ-OM-Sed scheme, as TOF of 10-yrs is possible with this scheme. If the constraint (1) needs to be reduced to 9 km/s, using EJSed is best since it takes 15.1 years to reach Sedna (Table 10). The trajectories of the spacecraft flight using EJSed and EJ-OM-Sed schemes are given in Fig 3 and Fig 16.

8. **Possible expansion of space mission to Sedna**

The global practice of maximising the scientific output through expanding the research program of space missions is well known. For example, during the NEAR [68,69] mission to the asteroid (433) Eros, it also approached the asteroid (253) Matilda, making it possible to determine its mass and obtain photos of its surface. There are also known examples of Galileo [70] (approaching (243) Ida and (951) Gaspra), Cassini–Huygens ((2685) Masursky) [71], Ulysses (passing the gas tail of a few comets), and New Horizons [72] (TNO (486958) Arrokot flyby) missions, etc. Taking as an example the mentioned missions, we carry out a study of possible expansion of the flight scenario to Sedna.

One way of expanding the mission to Sedna may be to explore the planets and their satellites during gravity assist manoeuvres. Another way of expanding the mission scenario could be provided by including close encounters of small celestial bodies. In the paper [43], the main belt asteroids that could be encountered during the flight to Sedna were considered as potential additional targets. This paper proposes a different way of expansion: sending a small spacecraft to another TNO during the primary flight to Sedna. For that purpose, it is proposed that the spacecraft should consist of two modules: a vehicle to study Sedna and a small space probe, which would be separated from the main spacecraft during the last gravity assist manoeuvre and then directed to another TNO. Note that similar research of study two objects in one mission with spacecraft separation was proposed in articles [23,61] for studying classical KBOs.

The general optimization scheme considered above will not be significantly changed. However, there is a specific feature of the algorithm used in this section. Like in the previous case, the model of patched conic approximation is used. Impulses needed during the gravity assist are described in sec. 3.1. The new one added in this section is required for a separation of the probe from the main spacecraft and directing it to another TNO ($\Delta V_s$). The $\Delta V_s$ value was calculated using the method described in [65]. Note that for implementation of the method the separation of spacecrafts is considered to be on the SOI boundary of the last flyby planet.

Let us limit the magnitude of the separation impulse as follows:

$$\Delta V_s \leq 300 \; m/s \qquad (3)$$

No restrictions are imposed on the TOF to the second chosen TNO. It is assumed that the flight to another TNO is unpowered, i.e. without any additional manoeuvres. Nevertheless, the used optimisation process considers the manoeuvres of the probe inside the SOI. Hence all probe trajectory required powered gravity assist were removed in the final selection process.

A total of 5 suitable trans-Neptunian bodies were selected (Table 11), namely three extreme TNOs (2012 VP$_{113}$ ('Biden'), Leleākūhonua (former 2015 TG$_{387}$), 2013 SY$_{99}$) and two classical Edgeworth-Kuiper belt objects: (90482) Orcus and (20000) Varuna. The results for all flights satisfying constraints (1), (2) and (3) are presented in Table 12.

Table 11. Parameters of Sedna and the TNOs suitable for proposed mission expansion

| TNO | d*, km | Orbital elements** | | | | | | T, yrs |
| --- | --- | --- | --- | --- | --- | --- | --- | --- |
| | | q, AU | a, AU | e | i, deg | Ω, deg | ω, deg | |
| Sedna [8] | ~1000 | 76.341 | 499.469 | 0.847 | 11.931 | 144.208 | 311.130 | 11,162.77 |
| 2012 VP$_{113}$ [21] | ~600 | 80.412 | 266.462 | 0.698 | 24.089 | 90.720 | 293.659 | 4,349.73 |
| Leleākūhonua [4] | ~300 | 65.172 | 1,210.470 | 0.946 | 11.655 | 300.795 | 117.648 | 42,115.24 |
| 2013 SY$_{99}$[73] | ~250 | 50.080 | 792.110 | 0.936 | 4.216 | 29.526 | 31.718 | 22,293.92 |
| Orcus [7] | 761…917 | 30.155 | 39.110 | 0.228 | 20.584 | 268.725 | 72.701 | 244.60 |
| Varuna [74,75] | 859×453 | 40.265 | 42.724 | 0.057 | 17.209 | 97.350 | 263.222 | 279.27 |

* d is TNOs diameter, calculated assuming albedo taken in https://ssd.jpl.nasa.gov/sbdb.cgi#top; **Orbital elements are given w.r.t. to Ecliptic plane J2000 Epoch; Designation used in the table: q is the perihelion distance, a is the semi-major axis, i is the inclination, e is the eccentricity, Ω is the longitude of ascending node, ω is the argument of pericentre, T is the orbital period.

Table 12. Catalogue of selected schemes providing exploration of both Sedna and another TNO simultaneously

| Target TNOs | Scheme | Launch date, dd.mm.yyyy | TOF, yrs | $\Delta V_\Sigma$, km/s | Date of separation, dd.mm.yyyy | Pericentre height*, 10$^3$ km | TOF$_1$**, yrs | $\Delta V_s$***, km/s | $V_c$****, km/s |
| --- | --- | --- | --- | --- | --- | --- | --- | --- | --- |
| 2012 VP$_{113}$ | | | | | | | | | |
| | EJSed | | | | | | | | |
| | | 01.04.2032 | 18 | 7.90 | 24.06.2033 | 55.0 | 19.8 | 0.020 | 19.91 |
| | | 31.03.2032 | 20 | 7.42 | 24.07.2033 | 101.8 | 21.8 | 0.023 | 17.97 |
| | | 10.05.2033 | 18 | 8.22 | 27.06.2034 | 429.4 | 21.2 | 0.039 | 18.04 |
| | | 09.05.2033 | 20 | 7.87 | 21.07.2034 | 617.5 | 24.8 | 0.052 | 15.27 |
| | | 22.06.2034 | 18 | 9.82 | 31.05.2035 | 1421.2 | 23.1 | 0.167 | 16.24 |
| | | 20.06.2034 | 20 | 9.37 | 15.06.2035 | 1926.8 | 27.0 | 0.225 | 13.72 |
| | | 26.01.2039 | 20 | 9.87 | 13.08.2045 | 3.6 | 30.7 | 0.034 | 12.81 |
| | EJNSed | | | | | | | | |
| | | 19.12.2040 | 20 | 8.63 | 11.08.2043 | 43.6 | 14.1 | 0.012 | 20.82 |
| | | 19.01.2041 | 20 | 7.94 | 23.01.2045 | 51.0 | 13.4 | 0.014 | 21.87 |
| | | 28.02.2043 | 18 | 8.86 | 22.02.2045 | 52.6 | 13.5 | 0.015 | 21.76 |
| | | 28.02.2043 | 20 | 7.58 | 09.02.2045 | 87.8 | 15.6 | 0.019 | 18.73 |
| | | 06.04.2044 | 18 | 7.65 | 07.07.2045 | 69.7 | 14.4 | 0.017 | 20.25 |
| | | 04.04.2044 | 20 | 7.29 | 12.08.2045 | 110.6 | 16.6 | 0.022 | 17.56 |
| | EΔVEJSed | | | | | | | | |
| | | 27.02.2029 | 18 | 7.68 | 18.09.2033 | 244.9 | 26.5 | 0.01 | 14.45 |

| | | | | | | | | |
|---|---|---|---|---|---|---|---|---|
| | | 02.03.2029 | 20 | 6.36 | 29.06.2033 | 56.2 | 19.8 | 0.01 | 19.90 |
| | | 15.04.2030 | 18 | 8.07 | 04.06.2034 | 282.5 | 18.3 | 0.01 | 21.15 |
| | | 13.04.2030 | 20 | 6.92 | 17.06.2034 | 340.7 | 19.5 | 0.01 | 19.75 |
| | EΔVEJNSed | | | | | | | | |
| | | 03.05.2040 | 20 | 6.31 | 11.05.2045 | 40.82 | 10.6 | 0.013 | 23.41 |
| | | 09.05.2041 | 18 | 8.99 | 20.05.2045 | 29.06 | 12.5 | 0.01 | 25.36 |
| | | 05.05.2041 | 20 | 6.18 | 12.05.2045 | 54.84 | 11.6 | 0.015 | 21.67 |
| | | 17.06.2042 | 18 | 9.08 | 11.06.2045 | 35.58 | 13.5 | 0.012 | 24.08 |
| | | 13.06.2042 | 20 | 7.38 | 29.04.2046 | 62.28 | 12.2 | 0.016 | 20.58 |
| Leleākūhonua | | | | | | | | | |
| | EJSed | | | | | | | | |
| | | 01.04.2032 | 18 | 7.90 | 25.06.2033 | 757.0 | 19.5 | 0.176 | 16.03 |
| | | 31.03.2032 | 20 | 7.42 | 24.07.2033 | 1163.1 | 24.9 | 0.23 | 12.07 |
| | | 05.04.2044 | 18 | 7.85 | 22.07.2045 | 759.8 | 20.0 | 0.172 | 14.73 |
| | | 05.04.2044 | 20 | 7.39 | 01.08.2045 | 867.3 | 21.4 | 0.18 | 13.58 |
| | EΔVEJSed | | | | | | | | |
| | | 27.02.2029 | 18 | 7.68 | 18.09.2033 | 1867.3 | 50.9 | 0.300 | 5.1 |
| | | 03.03.2029 | 20 | 6.36 | 29.06.2033 | 753.5 | 19.5 | 0.200 | 16.0 |
| | | 16.01.2038 | 20 | 9.10 | 09.09.2045 | 3.6 | 68.6 | 0.100 | 3.7 |
| 2013 SY$_{99}$ | | | | | | | | | |
| | EJSed | | | | | | | | |
| | | 27.01.2030 | 18 | 8.90 | 16.04.2033 | 39.1 | 49.0 | 0.060 | 3.82 |
| | | 01.04.2032 | 18 | 7.90 | 24.06.2033 | 271.3 | 11.1 | 0.085 | 20.81 |
| | | 31.03.2032 | 20 | 7.42 | 24.07.2033 | 399.2 | 12.5 | 0.095 | 18.28 |
| | | 10.05.2033 | 18 | 8.22 | 27.06.2034 | 1,419.8 | 14.8 | 0.274 | 14.46 |
| | | 05.04.2044 | 18 | 7.85 | 21.07.2045 | 169.2 | 11.3 | 0.070 | 19.70 |
| | | 04.04.2044 | 20 | 7.39 | 01.08.2045 | 195.2 | 11.6 | 0.072 | 18.97 |
| | EΔVEJSed | | | | | | | | |
| | | 07.05.2030 | 18.0 | 7.31 | 01.08.2033 | 455.2 | 13.1 | 0.119 | 17.23 |
| | | 07.05.2030 | 20.0 | 6.70 | 25.07.2033 | 417.3 | 12.7 | 0.106 | 17.82 |
| | | 14.06.2031 | 20.0 | 7.17 | 24.06.2034 | 1,411.9 | 14.9 | 0.271 | 14.32 |
| | | 06.03.2040 | 20.0 | 7.77 | 01.09.2044 | 119.6 | 43.8 | 0.056 | 4.04 |
| | | 11.04.2041 | 20.0 | 8.72 | 03.10.2045 | 195.9 | 58.0 | 0.073 | 3.16 |
| | | 10.05.2042 | 18.0 | 7.26 | 11.10.2045 | 491.6 | 15.4 | 0.127 | 13.62 |
| | | 20.06.2043 | 18.0 | 7.61 | 09.06.2046 | 521.1 | 10.7 | 0.099 | 19.91 |
| | | 18.06.2043 | 20.0 | 6.99 | 09.07.2046 | 788.0 | 12.8 | 0.129 | 16.33 |
| | EΔVEJNSed | | | | | | | | |
| | | 21.09.2041 | 20 | 9.21 | 11.05.2051 | 136.5 | 4.5 | 0.037 | 21.30 |
| | | 14.05.2042 | 18 | 7.19 | 01.12.2050 | 108.4 | 4.1 | 0.034 | 23.71 |
| | | 12.05.2042 | 20 | 6.47 | 11.09.2051 | 145.6 | 4.7 | 0.038 | 20.56 |
| | | 21.06.2043 | 18 | 8.22 | 07.09.2051 | 114.6 | 4.3 | 0.034 | 22.62 |
| | | 20.06.2043 | 20 | 7.58 | 01.06.2052 | 152.4 | 4.9 | 0.038 | 19.62 |
| | EJNSed | | | | | | | | |
| | | 19.12.2040 | 20 | 8.63 | 12.06.2050 | 127.1 | 4.6 | 0.035 | 21.19 |
| | | 19.01.2041 | 20 | 7.94 | 21.01.2051 | 124.9 | 4.3 | 0.036 | 22.26 |
| Orcus | | | | | | | | | |
| | EΔVEJSed | | | | | | | | |
| | | 05.06.2029 | 18 | 9.06 | 06.07.2034 | 3.6 | 25.5 | 0.100 | 7.68 |
| | | 25.02.2029 | 20 | 7.21 | 07.10.2033 | 3.6 | 38.0 | 0.100 | 4.27 |
| | | 18.02.2030 | 18 | 7.33 | 14.10.2033 | 3.6 | 40.6 | 0.100 | 3.90 |
| | | 10.04.2030 | 20 | 7.23 | 16.07.2034 | 3.6 | 25.7 | 0.100 | 7.63 |
| | | 05.04.2031 | 20 | 7.40 | 15.07.2034 | 3.6 | 25.8 | 0.100 | 7.60 |
| Varuna | | | | | | | | | |
| | EJSed | | | | | | | | |
| | | 31.03.2032 | 18 | 7.92 | 14.07.2033 | 3.6 | 25.1 | 0.069 | 7.11 |
| | | 31.03.2032 | 20 | 7.42 | 24.07.2033 | 10.9 | 26.5 | 0.078 | 6.60 |
| | | 10.05.2033 | 18 | 8.22 | 27.06.2034 | 79.4 | 23.6 | 0.132 | 7.93 |
| | | 08.05.2033 | 20 | 7.87 | 21.07.2034 | 92.7 | 23.1 | 0.142 | 8.06 |
| | | 22.06.2034 | 18 | 9.82 | 31.05.2035 | 108.3 | 55.4 | 0.238 | 4.20 |
| | | 20.06.2034 | 20 | 9.37 | 15.06.2035 | 131.8 | 55.1 | 0.269 | 4.01 |
| | | 03.02.2038 | 20 | 9.82 | 26.12.2045 | 1,087.2 | 36.4 | 0.22 | 4.34 |
| | | 06.02.2039 | 20 | 9.16 | 04.01.2046 | 1,166.7 | 39.7 | 0.223 | 3.87 |
| | | 07.02.2040 | 18 | 9.80 | 01.12.2045 | 1,418.7 | 46.3 | 0.247 | 3.19 |
| | | 09.02.2040 | 20 | 8.66 | 03.01.2046 | 1,319.2 | 45.2 | 0.234 | 3.30 |
| | EΔVEJSed | | | | | | | | |
| | | 05.06.2029 | 18 | 8.25 | 24.04.2034 | 43.2 | 23.9 | 0.100 | 8.14 |
| | | 16.04.2029 | 20 | 7.08 | 10.06.2034 | 67.6 | 24.6 | 0.100 | 7.67 |
| | | 09.06.2030 | 18 | 7.83 | 06.05.2034 | 49.5 | 23.9 | 0.100 | 8.08 |
| | | 13.04.2030 | 20 | 6.92 | 17.06.2034 | 72.3 | 24.4 | 0.100 | 7.70 |

\* Height of the pericentre calculated for the probe trajectories after the probe separation  
\*\* TOF$_1$ is the time of flight after the probe separation to the TNO;  
\*\*\* ΔV$_s$ is the separation impulse at a distance of 80 mln. km from the flyby planet;  
\*\*\*\* V$_e$ is the flyby velocity near the second TNO.

For example, the flight to Sedna and the sednoid 2012 VP$_{113}$ ('Biden') using the EΔVEJSed scheme with the launch on 02.03.2029 and TOF (to Sedna) equal to 20-yr is considered. The orbital diagram is shown in Fig. 19; the corresponding spacecraft's trajectory parameters are shown in Table 13.

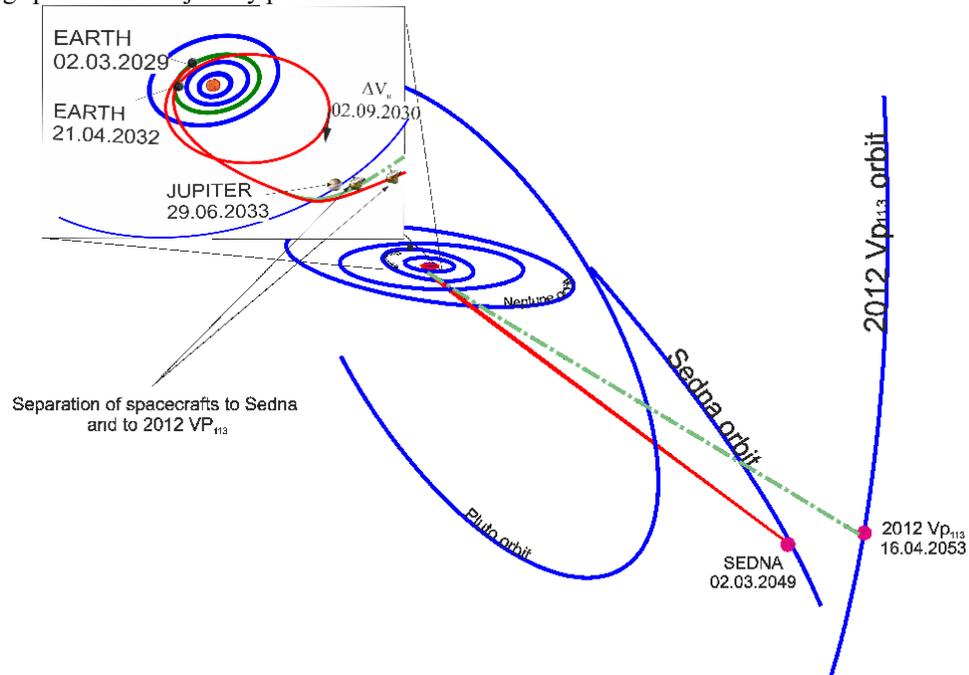

Fig. 19 Trajectory of spacecraft and probe flight to Sedna and 2012 VP$_{113}$ for launch on 02.03.2029 using EΔVEJSed scheme

Table 13. The spacecraft and probe trajectory parameters to Sedna and 2012 VP$_{113}$ using EΔVEJSed scheme for launch on 02.03.2029

| Celestial bodies/deep space manoeuvres | Dates of launch and flyby of celestial bodies, dd.mm.yyyy | Relative velocities near-Earth and of the flyby of celestial bodies, km/s | ΔV of launch, at aphelion and of the flyby of celestial bodies, km/s | Height of the initial orbit and flyby above celestial bodies, $10^3$ km |
|---|---|---|---|---|
| Earth | 02.03.2029 | 7.02 | 5.28 | 0.2 |
| ΔV$_α$ | 02.09.2030 | 10.64 | 0.60 | - |
| Earth | 21.04.2032 | 12.36 | 0.00 | 0.3 |
| Jupiter (to Sedna) | 29.06.2033 | 16.46* | 0.48 | 3.7 |
| Jupiter (to 2012 VP$_{113}$) | 29.06.2033 | 16.46* | 0 | 56.3 |
| Sedna (the spacecraft) | 02.03.2049 | 22.44 | - | 0.0** |
| 2012 VP$_{113}$ (the probe) | 16.04.2053 | 19.88 | - | 0.0** |

\* The value of incoming relative velocity is given.
\*\*Approach to the object is assumed to be at any small distance

## 9. Conclusions

In this paper, the flight schemes to Sedna, providing its fast approach for studying the object at the closest possible distance were analyzed. Both the direct flight scheme to the object and schemes including a series of gravity assists of planets were considered. It is demonstrated that a flight to Sedna can be performed almost each year beginning from 2029 under a constraint on the TOF (constraint (1)) with the ΔV$_Σ$ limit up to 12 km/s (constraint (2)). Moreover, the ΔV$_Σ$ limit (2) can be easily reduced to ~9 km/s.

The direct flight to Sedna can be performed each year because of the synodic period of Earth and Sedna. Flight with considered constraint on TOF of 20-yr requires launch ΔV of at least

14 km/s. Thus, a direct flight to Sedna is hardly possible with the existing technical means because the flight with duration of 20 years or less would require usage of super heavy launch vehicles.

The best of the EJSed, EJNSed, EΔVEJSed, EΔVEJNSed, EVEΔVEJSed (EVEEJSed), EJ-OM-Sed flight schemes proved to be the EΔVEJNSed scheme for TOF=20-yr, providing the minimum $\Delta V_\Sigma$ = 6.13 km/s. In that case, 490 kg can be delivered by a light class Soyuz 2.1.b rocket and 2,070 kg by a medium-class Proton-M rocket. If the main criterion of choosing the flight scheme is to reach Sedna as soon as possible, and the $\Delta V_\Sigma$ limit (1) must be met, the EJ-OM-Sed scheme is best, allowing a ~100 kg payload mass delivery to Sedna by the SLS. However, this scheme would require a close approach to the Sun of up to 2 solar radii, so using the classic and well known EJSed scheme seems most appropriate. It takes 15.1 years to reach Sedna at the cost of $\Delta V_\Sigma$ = 9.01 km/s using the EJSed scheme, and a 295.5 kg payload would be delivered to Sedna by Proton-M in this case.

The estimation of the final mass of the spacecraft delivered to Sedna shows that the results of $\Delta V_\Sigma$ for the schemes considered in the paper allow to easily obtain from 50 to 12,000 kg on the flyby trajectory near Sedna depending on the launch vehicle used. The maximum payload mass for all investigated schemes is achieved at TOF=20-yr. B, Proton-M, Delta-IV Heavy and SLS, respectively. For the EJSed and EJNSed schemes the maximum payloads using Soyuz 2.1.b and Proton-M rockets will be ~300 kg and ~1,200 kg while using Delta-IV Heavy and SLS to Sedna provides ~1,300 kg and ~4,500 kg respectively. The ΔVEGA manoeuvre schemes (EΔVEJSed, EΔVEJNSed and EVEΔVEJSed (EVEEJSed)) using Soyuz 2.1 can deliver maximum 500 kg, 2000 kg, 2,000 kg and 12,000 kg. The EJ-OM-Sed scheme would achieve the maximum payload mass of about 300 kg and 1,000 kg using Delta IV Heavy and SLS rockets.

As a potential expansion of the mission to Sedna, the scenario of a two-target mission was considered. One of the targets of this mission is Sedna, and the others are the TNOs that can be reached without an additional increase of $\Delta V_\Sigma$. Five TNOs have been found to which a flight under such conditions is possible. Three of these objects are extreme TNOs (2012 VP$_{113}$, 2013 SY99, Leleākūhonua) with an orbital period of about 4, 22 and 44 thousand years correspondingly. The others are classic KBOs: Orcus and Varuna. The best results were obtained for 2012 VP$_{113}$, as passive flight is achieved by almost any of the schemes considered (see Table 12).

Summarising all of the above, we would like to emphasize that despite the distance of Sedna from the Earth, at the current level of rocket and space technology, there are hardly technical difficulties in exploring Sedna even from the aspect of flyby trajectory. Yet the exploration of such an object could be the first step for mankind to interstellar space.

**Acknowledgments**

The author is sincerely grateful to Dr Alexander Sukhanov and Dr Konstantin Fedyaev, leading mathematicians of the Space Research Institute of the Russian Academy of Sciences, and Dr Natan Eismont, the leading researcher of the Space Research Institute of the Russian Academy of Sciences, Dr Vsevolod Koryanov and to my colleague Andrey Belyaev for their essential help during the paper's preparation and their help with the English translation and proof-reading.

Also, author want to sincerely thank the participants of the Seminar on mechanics, control and informatics dedicated to optimal flight to Sedna6. The meeting led by Prof. R.R. Nazirov was held in the Space Research Institute (IKI) of the Russian Academy of Sciences. The seminar participants' comments helped to improve obtained results, which subsequently was included in this paper.**Appendix A**

In this appendix, the trajectory catalogues for the flight schemes to Sedna discussed in this paper are compiled. Tables A.1-A.2 and A.5 contain data for the flight schemes without a gravity manoeuvre near Neptune. The catalogues are built for launch dates from 2029 to 2042, i.e. for one orbital period of Jupiter. Tables A.3 and A.4 are derived for schemes with gravity assist of Neptune for launch dates after 2039. The trajectories in all the below tables satisfy constraints (1) and (2) simultaneously.

Table A.1. Catalogue of selected trajectories to Sedna by the EJSed scheme

| Optimal launch date, dd.mm.yyyy | Launch window*, dd.mm | TOF, yrs | Jupiter flyby date, dd.mm.yyyy | Height of Jupiter flyby, $10^3$ km | $\Delta V_\Sigma$, km/s | $V_f$, km/s |
|---|---|---|---|---|---|---|

---

[6] The Seminar's web page http://iki.rssi.ru/seminar/20210611/e_abstract.php

| Optimal launch date, dd.mm.yyyy | Launch window*, dd.mm | TOF, yrs | Jupiter flyby date, dd.mm.yyyy | Height of Jupiter flyby, $10^3$ km | $\Delta V_\Sigma$, km/s | $V_f$, km/s |
|---|---|---|---|---|---|---|
| 28.01.2029 | 25.01-05.02 | 16 | 25.07.2033 | 3.6 | 10.77 | 33.01 |
| 31.01.2029 | 24.01-09.02 | 18 | 19.09.2033 | 3.6 | 9.25 | 28.08 |
| 03.02.2029 | 23.01-11.02 | 20 | 21.10.2033 | 3.6 | 8.34 | 24.31 |
| 27.01.2030 | 17.01-05.02 | 16 | 18.03.2033 | 3.6 | 10.56 | 29.72 |
| 27.01.2030 | 16.01-05.02 | 18 | 16.04.2033 | 3.6 | 8.90 | 25.60 |
| 01.02.2030 | 22.01-03.02 | 20 | 09.07.2033 | 3.6 | 8.10 | 22.51 |
| 17.02.2031 | 08.02-25.02 | 18 | 16.10.2033 | 3.6 | 10.55 | 24.24 |
| 16.02.2031 | 07.02-26.02 | 20 | 09.10.2033 | 3.6 | 9.40 | 21.25 |
| 30.03.2032 | 08.02-26.02 | 14 | 14.12.2033 | 3.6 | 9.95 | 30.42 |
| 31.03.2032 | 20.03-10.04 | 16 | 28.07.2033 | 3.6 | 8.57 | 25.57 |
| 01.04.2032 | 21.03-11.04 | 18 | 25.06.2033 | 14.1 | 7.90 | 22.24 |
| 31.03.2032 | 21.03-10.04 | 20 | 24.07.2033 | 51.2 | 7.42 | 19.73 |
| 18.05.2033 | 02.05-21.05 | 12 | 13.03.2034 | 30.0 | 11.44 | 33.29 |
| 15.05.2033 | 30.04-20.05 | 14 | 23.04.2034 | 92.8 | 9.76 | 28.22 |
| 12.05.2033 | 02.05-21.05 | 16 | 29.05.2034 | 176.3 | 8.80 | 24.39 |
| 10.05.2033 | 30.04-20.05 | 18 | 27.06.2034 | 277.9 | 8.22 | 21.40 |
| 08.05.2033 | 30.04-19.05 | 20 | 21.07.2034 | 392.6 | 7.87 | 19.00 |
| 25.06.2034 | 14.06-02.07 | 14 | 20.04.2035 | 408.3 | 11.49 | 27.11 |
| 23.06.2034 | 14.06-02.07 | 16 | 12.05.2035 | 596.4 | 10.47 | 23.46 |
| 22.06.2034 | 14.06-02.07 | 18 | 31.05.2035 | 805.9 | 9.82 | 20.60 |
| 20.06.2034 | 14.06-02.07 | 20 | 15.06.2035 | 1,027.9 | 9.37 | 18.30 |
| 02.08.2035 | 27.07-13.08 | 20 | 25.05.2036 | 1,960.3 | 11.82 | 17.89 |
| 31.01.2037 | 21.01-10.02 | 20 | 10.12.2045 | 3.6 | 10.68 | 32.82 |
| 03.02.2038 | 22.10-07.11 | 20 | 26.12.2045 | 3.6 | 9.82 | 30.05 |
| 03.02.2039 | 25.01-12.02 | 18 | 24.11.2045 | 3.6 | 10.65 | 32.71 |
| 06.02.2039 | 28.01-17.02 | 20 | 04.01.2046 | 3.6 | 9.16 | 27.67 |
| 03.02.2040 | 24.01-12.02 | 16 | 29.12.2043 | 3.6 | 11.76 | 35.74 |
| 07.02.2040 | 28.01-19.02 | 18 | 15.02.2045 | 3.6 | 9.80 | 29.92 |
| 10.02.2040 | 27.01-19.02 | 20 | 04.07.2045 | 3.6 | 8.66 | 25.59 |
| 06.02.2041 | 27.01-15.02 | 16 | 27.09.2045 | 3.6 | 10.63 | 32.39 |
| 09.02.2041 | 30.01-18.02 | 18 | 18.11.2045 | 3.6 | 9.15 | 27.48 |
| 11.02.2041 | 01.02-23.02 | 20 | 17.12.2045 | 3.6 | 8.27 | 23.74 |

*Launch window calculated considering about +0.25 km/s increase in $\Delta V_\Sigma$ on the boundaries of window.*

Table A.2. Catalogue of selected trajectories to Sedna by the EΔVEJSed scheme

| Optimal launch date, dd.mm.yyyy | Launch window*, dd.mm | TOF, yrs | Duration of E-E trajectory part, yrs | Jupiter flyby date, dd.mm.yyyy | Height of Jupiter flyby, $10^3$ km | $\Delta V_\Sigma$, km/s | $V_f$, km/s |
|---|---|---|---|---|---|---|---|
| 27.02.2029 | 16.02-20.03 | 16 | 3 | 03.12.2033 | 3.6 | 10.46 | 33.41 |
| 26.02.2029 | 15.02-19.03 | 18 | 3 | 18.09.2033 | 3.6 | 7.68 | 27.93 |
| 02.03.2029 | 18.02-22.03 | 20 | 3 | 29.06.2033 | 3.6 | 6.36 | 23.91 |
| 11.06.2030 | 27.05-27.06 | 14 | 3 | 23.03.2034 | 21.8 | 11.58 | 36.50 |
| 10.06.2030 | 26.05-26.06 | 16 | 3 | 30.03.2034 | 57.6 | 9.30 | 30.36 |
| 20.02.2030 | 01.02-11.03 | 18 | 2 | 20.09.2033 | 15.1 | 7.23 | 25.94 |
| 20.02.2030 | 01.02-11.03 | 20 | 2 | 16.09.2033 | 69.4 | 6.59 | 19.92 |
| 22.04.2031 | 02.04-07.05 | 14 | 2 | 19.03.2034 | 31.3 | 10.84 | 33.56 |
| 16.04.2031 | 22.03-26.04 | 18 | 2 | 30.05.2034 | 90.84 | 7.86 | 24.54 |
| 05.04.2031 | 18.03-22.04 | 20 | 2 | 29.06.2034 | 274.1 | 7.17 | 21.53 |
| 29.05.2032 | 14.05-18.06 | 16 | 2 | 21.04.2035 | 404.3 | 10.87 | 27.28 |
| 25.05.2032 | 11.05-15.06 | 18 | 2 | 14.05.2035 | 591.8 | 9.75 | 23.59 |
| 22.05.2032 | 08.05-13.06 | 20 | 2 | 02.06.2035 | 801.4 | 9.01 | 20.71 |
| 09.03.2037 | 13.02-21.03 | 20 | 3 | 09.11.2045 | 3.6 | 9.28 | 32.35 |
| 13.03.2038 | 18.02-26.03 | 20 | 3 | 29.10.2045 | 3.6 | 8.57 | 29.54 |
| 11.03.2039 | 16.02-25.03 | 20 | 3 | 16.07.2045 | 3.6 | 8.15 | 26.89 |
| 12.03.2040 | 17.02-26.03 | 20 | 3 | 05.10.2045 | 3.6 | 7.53 | 25.17 |
| 11.06.2041 | 29.05-24.06 | 16 | 3 | 26.04.2046 | 39.9 | 9.72 | 32.44 |
| 11.06.2041 | 29.05-24.06 | 18 | 3 | 29.04.2046 | 77.3 | 7.95 | 27.31 |

| Optimal launch date, dd.mm.yyyy | Launch window*, dd.mm | TOF, yrs | Duration of E-E trajectory part, yrs | Jupiter flyby date, dd.mm.yyyy | Height of Jupiter flyby, $10^3$ km | $\Delta V_\Sigma$, km/s | $V_f$, km/s |
|---|---|---|---|---|---|---|---|
| 05.03.2041 | 25.02-21.03 | 20 | 3 | 19.08.2045 | 3.6 | 6.52 | 23.36 |
| 14.06.2042 | 26.05-22.06 | 14 | 3 | 14.04.2046 | 19.4 | 11.14 | 35.65 |
| 24.02.2042 | 13.02-15.03 | 16 | 3 | 08.01.2046 | 3.6 | 8.97 | 29.96 |
| 13.06.2042 | 27.05-21.06 | 18 | 3 | 19.05.2046 | 115.2 | 7.59 | 25.34 |
| 17.04.2042 | 02.04-27.04 | 20 | 3 | 01.07.2046 | 203.6 | 6.69 | 22.32 |

*Launch window calculated considering about +0.25 km/s increase in $\Delta V_\Sigma$ on the boundaries of window.*

Table A.3. Catalogue of selected trajectories to Sedna by the E$\Delta$VEJNSed scheme

| Optimal launch date, dd.mm.yyyy | Launch window*, dd.mm | TOF, yrs | Duration of E-E trajectory part, yrs | Jupiter flyby date, dd.mm.yyyy | Height of Jupiter flyby, $10^3$ km | Height of Neptune flyby, $10^3$ km | $\Delta V_\Sigma$, km/s | $V_f$, km/s |
|---|---|---|---|---|---|---|---|---|
| 06.05.2040 | 30.04-10.05 | 18 | 3 | 11.05.2045 | 27.9 | 1.2 | 11.42 | 30.18 |
| 03.05.2040 | 23.04-08.05 | 20 | 3 | 20.05.2045 | 63.5 | 1.2 | 6.31 | 25.99 |
| 09.05.2041 | 02.05-13.05 | 18 | 3 | 11.05.2045 | 41.8 | 1.2 | 8.99 | 27.90 |
| 08.03.2041 | 01.03-15.03 | 20 | 3 | 24.06.2045 | 100.4 | 4.6 | 6.13 | 24.31 |
| 30.04.2042 | 22.04-07.05 | 17 | 3 | 21.04.2046 | 306.6 | 1.2 | 12.01 | 29.08 |
| 27.04.2042 | 18.04-05.05 | 18 | 3 | 07.05.2046 | 307.4 | 1.2 | 9.50 | 26.98 |
| 20.04.2042 | 12.04-28.04 | 20 | 3 | 04.06.2046 | 461.9 | 6.7 | 7.43 | 23.52 |

*Launch window calculated considering about +0.25 km/s increase in $\Delta V_\Sigma$ on the boundaries of window.*

Table A.4. Catalogue of selected trajectories to Sedna by the EJNSed scheme

| Optimal launch date, dd.mm.yyyy | Launch window*, dd.mm | TOF, yrs | Jupiter flyby date, dd.mm.yyyy | Height of Jupiter flyby, $10^3$ km | Height of Neptune flyby, $10^3$ km | $\Delta V_\Sigma$, km/s | $V_f$, km/s |
|---|---|---|---|---|---|---|---|
| 24.12.2039 | 22.12-12.01/40 | 20 | 23.10.2044 | 74.5 | 1.2 | 9.84 | 26.12 |
| 19.12.2040 | 12.12-12.01/41 | 20 | 11.08.2043 | 3.7 | 1.2 | 8.63 | 23.41 |
| 17.01.2041 | 08.01-27.01 | 18 | 10.12.2044 | 3.6 | 1.2 | 10.99 | 28.09 |
| 19.01.2041 | 09.01-27.01 | 20 | 23.01.2045 | 3.6 | 4.5 | 7.94 | 24.49 |
| 24.01.2042 | 15.01-12.02 | 18 | 17.12.2044 | 3.6 | 1.2 | 11.23 | 26.11 |
| 24.01.2042 | 15.01-12.02 | 20 | 18.12.2044 | 3.6 | 9.2 | 10.01 | 22.86 |
| 28.02.2043 | 20.02-11.03 | 18 | 22.02.2045 | 3.6 | 4.9 | 8.86 | 24.37 |
| 27.02.2043 | 18.02-09.03 | 20 | 08.02.2045 | 3.6 | 14.8 | 7.58 | 21.45 |
| 11.04.2044 | 30.03-18.04 | 15 | 03.05.2045 | 38.7 | 1.2 | 10.68 | 28.04 |
| 09.04.2044 | 30.03-18.04 | 16 | 27.05.2045 | 67.5 | 1.2 | 8.40 | 26.13 |
| 06.04.2044 | 27.03-16.04 | 18 | 07.07.2045 | 140.2 | 9.8 | 7.65 | 22.93 |
| 04.04.2044 | 25.03-14.04 | 20 | 12.08.2045 | 232.4 | 20.2 | 7.29 | 20.34 |
| 20.05.2045 | 08.05-27.05 | 15 | 04.05.2046 | 311.7 | 2.8 | 10.56 | 26.84 |
| 16.05.2045 | 06.05-26.05 | 18 | 12.06.2046 | 556.6 | 11.8 | 8.59 | 21.95 |
| 15.05.2045 | 05.05-25.05 | 20 | 02.07.2046 | 739.9 | 22.4 | 8.22 | 19.49 |
| 30.06.2046 | 21.06-09.07 | 16 | 24.04.2047 | 1,945.6 | 2.3 | 11.56 | 24.09 |
| 29.06.2046 | 22.06-08.07 | 18 | 10.05.2047 | 2,839.9 | 10.6 | 10.79 | 21.17 |
| 28.06.2046 | 23.06-06.07 | 20 | 24.05.2047 | 3,960.8 | 20.4 | 10.25 | 18.82 |

*Launch window calculated considering about +0.25 km/s increase in $\Delta V_\Sigma$ on the boundaries of window.*

Table A.5. Catalogue of selected trajectories to Sedna by the EVEEJSed and EVE$\Delta$VEJSed

| Optimal launch date, dd.mm.yyyy | Launch window*, dd.mm | TOF, yrs | Jupiter flyby date, dd.mm.yyyy | Height of Jupiter flyby, $10^3$ km | $\Delta V_\Sigma$, km/s | $V_f$, km/s |
|---|---|---|---|---|---|---|
| | | | EVEEJSed | | | |
| 16.11.2029 | 30.10-25.11 | 14 | 09.04.2034 | 23.0 | 11.77 | 38.92 |
| 10.10.2029 | 15.09-18.10 | 17 | 14.04.2034 | 52.2 | 8.21 | 29.60 |
| 07.10.2029 | 15.09-18.10 | 18 | 06.05.2034 | 72.6 | 7.42 | 27.39 |
| 10.11.2029 | 18.10-23.11 | 19 | 25.05.2034 | 108.7 | 6.72 | 25.28 |
| 09.11.2029 | 19.10-23.11 | 20 | 08.06.2034 | 201.0 | 6.27 | 23.60 |

| | | EVEΔVEJSed | | | | |
|---|---|---|---|---|---|---|
| 16.11.2029 | 30.10-25.11 | 14 | 09.04.2034 | 23.0 | 11.77 | 38.92 |
| 10.10.2029 | 15.09-18.10 | 17 | 14.04.2034 | 52.2 | 8.21 | 29.60 |
| 07.10.2029 | 15.09-18.10 | 18 | 06.05.2034 | 72.6 | 7.42 | 27.39 |
| 10.11.2029 | 18.10-23.11 | 19 | 25.05.2034 | 108.7 | 6.72 | 25.28 |
| 09.11.2029 | 19.10-23.11 | 20 | 08.06.2034 | 201.0 | 6.27 | 23.60 |

*\* Launch window calculated considering about +0.25 km/s increase in $\Delta V_\Sigma$ on the boundaries of window.*

Table A.6. Catalogue of selected trajectories to Sedna by the EJ-OM-Sed scheme

| Optimal launch date, dd.mm.yyyy | TOF, yrs | Jupiter flyby date, dd.mm.yyyy | Height of Jupiter flyby, $10^3$ km | $\Delta V_\Sigma$, km/s | $V_f$, km/s | $r_{sun}$*, $10^6$ km |
|---|---|---|---|---|---|---|
| 15.05.2029 | 18 | 07.02.2037 | 951.1 | 11.24 | 44.83 | 2.3 |
| 11.05.2029 | 20 | 28.12.2036 | 859.2 | 10.32 | 35.75 | 2.4 |
| 16.05.2030 | 18 | 14.01.2037 | 973.7 | 11.00 | 40.04 | 2.3 |
| 15.05.2030 | 20 | 09.01.2037 | 966.5 | 10.49 | 32.64 | 2.6 |
| 22.05.2031 | 16 | 06.02.2037 | 1,220.0 | 12.00 | 45.31 | 2.3 |
| 18.05.2031 | 18 | 24.12.2036 | 1,089.3 | 11.10 | 36.12 | 2.5 |
| 19.05.2031 | 20 | 10.01.2037 | 1,109.9 | 10.80 | 30.04 | 3.5 |
| 24.05.2032 | 16 | 16.01.2037 | 1,521.7 | 12.15 | 40.52 | 2.3 |
| 24.05.2032 | 18 | 19.01.2037 | 1,480.9 | 11.67 | 33.10 | 3.1 |
| 26.05.2032 | 20 | 07.02.2037 | 1,446.2 | 11.37 | 27.97 | 4.7 |
| 24.07.2035 | 10 | 06.10.2036 | 178.4 | 11.95 | 51.35 | 2.3 |
| 23.07.2035 | 12 | 20.10.2036 | 308.8 | 10.53 | 41.21 | 2.3 |
| 23.07.2035 | 14 | 30.10.2036 | 437.9 | 9.83 | 34.22 | 2.3 |
| 22.07.2035 | 16 | 11.11.2036 | 390.0 | 9.77 | 29.53 | 2.4 |
| 22.07.2035 | 17 | 30.11.2036 | 404.4 | 9.84 | 25.23 | 3.1 |
| 21.07.2035 | 19 | 14.12.2036 | 423.9 | 9.90 | 21.96 | 3.7 |
| 29.08.2036 | 14 | 17.10.2037 | 3.7 | 12.25 | 31.41 | 11.1 |
| 29.08.2036 | 16 | 18.10.2037 | 3.6 | 11.58 | 26.59 | 13.0 |
| 29.08.2036 | 18 | 23.10.2037 | 7.3 | 11.09 | 23.02 | 15.1 |
| 29.08.2036 | 20 | 31.10.2037 | 16.3 | 10.72 | 20.27 | 17.1 |
| 28.10.2038 | 20 | 24.11.2040 | 3.6 | 11.78 | 20.94 | 2.3 |
| 28.11.2039 | 20 | 28.01.2042 | 3.6 | 11.01 | 21.13 | 2.3 |
| 24.05.2040 | 18 | 18.04.2049 | 975.2 | 11.81 | 50.35 | 2.3 |
| 20.05.2040 | 20 | 17.03.2049 | 839.7 | 10.36 | 39.30 | 2.3 |

*\* perihelion distance of the spacecraft*

.